\documentclass[amsart,a4paper,preprint,unsortedaddress]{revtex4-1}
\usepackage{geometry} % see geometry.pdf on how to lay out the page. There's lots.
\usepackage{enumerate}
\usepackage{amsmath}
\usepackage{amssymb}
\usepackage[dvips]{graphicx}
\usepackage{color}
\usepackage{listings}
\usepackage{tabularx}
\usepackage{gensymb}
\usepackage[]{algorithm2e}
\usepackage[footnotesize]{subfigure}
\usepackage{float}
\newcommand{\redc}[1]{{\color{black} #1}}    % xxx

\begin{document}
\title{Computing Long Timescale Biomolecular Dynamics using Quasi-Stationary Distribution Kinetic Monte Carlo (QSD-KMC)}
\author{Animesh Agarwal}
%\affiliation{Theoretical Biology and Biophysics group, Los Alamos National Laboratory, New Mexico 87544, United States}
%\affiliation{Theoretical Biology and Biophysics group, Los Alamos National Laboratory, New Mexico 87544, United States}
\author{Nicolas W. Hengartner}
%\affiliation{Theoretical Biology and Biophysics group, Los Alamos National Laboratory, New Mexico 87544, United States}
\author{S. Gnanakaran}
\email{gnana@lanl.gov}
\affiliation{Theoretical Biology and Biophysics group, Los Alamos National Laboratory, New Mexico 87544, United States}
\author{Arthur F. Voter} 
\email{afv@lanl.gov}
\affiliation{Theoretical Division T-1, Los Alamos National Laboratory, New Mexico 87544, United States}

\begin{abstract}{\scriptsize 
It is a challenge to obtain an accurate model of the state-to-state 
dynamics of a complex biological system from molecular dynamics (MD) 
simulations. In recent years, Markov 
State Models have gained immense popularity for computing state-to-state 
dynamics from a pool of short MD simulations. However, the assumption 
that the underlying dynamics on the reduced space is Markovian induces 
a systematic bias in the model, especially in biomolecular systems 
with complicated energy landscapes. To address this problem, we have 
devised a new approach we call quasi-stationary distribution kinetic Monte 
Carlo (QSD-KMC) that gives accurate long time state-to-state evolution while retaining 
the entire time resolution even when the dynamics is highly non-Markovian.  
The proposed method is a kinetic Monte Carlo approach that takes advantage 
of two concepts: (i) the quasi-stationary 
distribution, the distribution that results when a trajectory remains 
in one state for a long time (the dephasing time), such that the 
next escape is Markovian, and (ii) dynamical corrections theory, 
which properly accounts for the correlated events that occur as a 
trajectory passes from state to state before it settles again. In practice, 
this is achieved by specifying, for each escape, the intermediate 
states and the final state that has resulted from the escape. 
Implementation of QSD-KMC imposes 
stricter requirements on the lengths of the trajectories than in 
a Markov State Model approach, as the trajectories must be long enough 
to dephase.  However, the QSD-KMC model produces state-to-state trajectories 
that are statistically indistinguishable from an MD trajectory mapped 
onto the discrete set of states, for an arbitrary choice of state 
decomposition. Furthermore, the aforementioned concepts can 
be used to construct a Monte Carlo approach 
to optimize the state boundaries regardless  of the initial choice of 
states. \redc{We demonstrate the QSD-KMC method on two one-dimensional model 
systems, one of which is a driven nonequilibrium system,  and on two 
well-characterized biomolecular systems. } \par}

\end{abstract}
%%% BEGIN DOCUMENT
\maketitle

\section{Introduction}

%\medskip
%{\bf may change the name from QSD-KMC to QSD-KMC. I wish I could get 
%both concepts into the name, but it would be too unwieldy to say DC-QSD-KMC}
%\medskip

Molecular dynamics (MD) is a powerful tool for probing complex processes 
in biological systems, such as protein folding~\cite{folding1, folding2}, 
protein-ligand binding~\cite{ligand1, ligand2}, and large scale conformational 
changes that lead to cellular processes~\cite{cell1}. 
However, directly accessing the time scales relevant for such biological 
processes with MD is challenging, as these processes typically occur 
on long time scales.  While the advent of specialized hardware for 
biological systems such as ANTON~\cite{anton} allows extension of 
the simulation time to millisecond timescales, 
such resources are not routinely available to researchers.  
This situation has motivated the development of models that can utilize 
information from a number of short MD simulations, which can be generated 
in parallel (e.g., using folding@home~\cite{home1, home2}), to say something 
about the characteristics of the longer timescale evolution.  These 
models take advantage of the fact that these systems typically spend 
a period of time in a single ``state'' of the system, occasionally 
making a transition to a new state. 
In particular, Markov state models 
(MSM's)~\cite{msm1, msm2, msm3, msm4, msmvalid},
which have become popular over the past decade, offer this type of approach.
In MSM's, the full configurational space is mapped onto a reduced 
space which is discretized into $n$ discrete states (``microstates'') and the long-time 
kinetics is modeled via an $n\times n$ transition matrix where an 
element $i,j$ of this matrix represents the probability of being 
in state $j$ at time $t$ (the ``lag time") given that the system 
was in state $i$ at time $t=0$.  From this transition matrix  one 
can get information on the long timescale dynamics of the system.  
However, the assumption that the underlying dynamics on the reduced 
space is Markovian induces a systematic bias in the model~\cite{msmvalid}.  

This error in MSM due to non-Markovian behavior can be reduced either 
by (1) making the spatial discretization finer, which depends on 
the choice of the input coordinates and clustering method used or 
(2) decreasing the time resolution of the model by increasing 
the lag time, although this may prohibit investigation of faster processes 
that are relevant to the problem. Thus, if the state space is not 
optimal or the transition probabilities are calculated using a short 
lag time, the MSM's may give incorrect results. Recently, aiming 
at this issue, Zuckermann and co-workers formulated non-Markovian 
estimators~\cite{zuc1, zuc2}, which give a more accurate description 
of the dynamics compared to MSM's. This analysis is based on the 
inclusion of a ``history," which is available in every MD trajectory.  
The method does not rely on the fine details of the states used and 
works well even if a fine time resolution is used.  

Another class of methods, which has been mostly used for studying 
material science systems~\cite{kmcmat1, kmcmat2, kmcmat3, kmcmat4}, 
is the kinetic Monte Carlo (KMC)~\cite{kmc1} approach, which builds 
on the fact that for most material systems the dynamical evolution 
consists of stochastic jumps from one metastable state to another, where 
the system stays in a state for a sufficiently long-time to lose its 
memory before making a transition to the next state. Thus, if the 
underlying dynamics is Markovian, this characteristic can be exploited 
to evolve the system from one state to another over long times. However, 
in biological systems the energy landscapes are comprised of highly 
populated metastable states connected via lightly populated intermediate 
states, which gives rise to multiple fast processes and correlated 
events (such as recrossings at the dividing surface between the states); 
in such scenarios the underlying dynamics does not exhibit ideal 
Markovian behavior.  

In this work we propose a new approach, quasi-stationary distribution KMC 
(QSD-KMC), that gives accurate long-time statistics without compromising 
the time resolution of the system even when the evolution of the 
system is complex in the way the trajectory jumps from one state 
to another, such that the system cannot be assumed to be Markovian.  
There are two key components in this KMC-based approach: (1) Calculation 
of the total escape rate out of a state; this is accomplished using 
the quasi-stationary distribution (QSD) concept (\cite{qsd1, qsd2, 
qsd3, qsd4}), which allows direct computation of a unique value for 
this rate once the correlation time, or {\it dephasing time}, is known.  
(2) Treatment of correlated events; the effect of correlated events 
is included by specifying, for each KMC escape, the intermediate 
states and final state resulting from the escape.  

With these two components, the QSD-KMC method produces, from a compact 
representation, state-to-state trajectories that are statistically indistinguishable 
from the underlying MD trajectories, for any choice of state decomposition.  
Moreover, we believe that the qualitative interpretation afforded 
by this approach, in the form of QSD escape rates and an understanding 
of the correlated events, is valuable even in situations where it 
is feasible to generate a single long MD trajectory.
The concepts in the QSD-KMC approach can also be used to design 
a Monte Carlo method for optimization of the state boundaries. Searching 
for state boundaries that minimize the total correlated-event time 
leads to states that are, in some sense, maximally metastable.  
We note here that the enforcement of QSD based state-to-state dynamics
adds some contraints on the length of the trajectories
that we use to construct the QSD-KMC model, unlike the case
with the MSM trajectories.
%We note here that unlike the MSM trajectories, there are some constraints 
%on the length of the trajectories that we use to construct the QSD-KMC model. 
\redc{To demonstrate the robustness of the QSD-KMC approach, we apply the method 
to two one-dimensional model systems, one equilibrium and one a nonequilibrium driven system,
and to two different biomolecular systems with varying conformational 
characteristics: dialanine and villin headpiece.} For the biomolecular 
systems, we first use state boundaries obtained via Perron Cluster-Cluster 
Analysis (PCCA)~\cite{pccaschuette} of the dominant eigenvectors 
obtained via diagonalization of the MSM transition matrix; for these 
states, we demonstrate that the state-to-state evolution characteristics 
from QSD-KMC match those of the underlying MD trajectories.  We then 
demonstrate that minimizing the total correlated-event time provides 
a viable approach for generating good state definitions, even starting 
from a simple, unphysical rectangular lumping of microstates into 
contiguous macrostates.

%{\bf Will need to update this, as we are adding a section and maybe changing the name of another:}
\redc{The paper is organized as follows: we begin 
Section~\ref{theory} with a brief overview of the QSD-KMC approach.  We then 
discuss the standard KMC approach, the quasi-stationary distribution, 
dynamical corrections theory in the context of both transition state 
theory and the QSD, the logical connection between QSD-KMC and parallel 
trajectory splicing (an accelerated molecular dynamics method), and 
the bootstrap interpretation of QSD-KMC.  In Section~\ref{technical}, we describe 
the technical aspects of the QSD-KMC method, such as calculation of 
rate constants, the estimation of dephasing time, the procedure for generating
short trajectories for constructing the QSD-KMC model, and finally lay out the 
algorithmic details. In Section~\ref{optimization}, we discuss how minimizing the correlated-event time
can be employed to optimize the state boundaries. In Section~\ref{MSMmethod}, we briefly
describe the procedure for comparing the results from MSMs and QSD-KMC.
Finally, in Section~\ref{discussion}, we first demonstrate the important 
concepts of QSD-KMC on two one-dimensional test systems and then
show the robustness of QSD-KMC in applications to the two biomolecular systems. } 

\section{Theory} \label{theory}

In this section, after a brief overview of the QSD-KMC method, we 
develop its theoretical underpinnings.  

\subsection{Brief Overview of the QSD-KMC method} \label{briefoverview}

The overall goal of the QSD-KMC method is to provide an accurate, 
compact model of the state-to-state dynamics of a complex biological 
system.  The accuracy should be maintained regardless of whether 
the states are defined in a way that gives naturally Markovian behavior, 
or whether such a definition can even be found.  The model, which 
is built from the information contained in a set of short molecular 
dynamics trajectories that probe the states of the system, consists 
of two main parts: 1) a set of first-order rate constants that are used 
in a KMC procedure to generate the next escape from the current state 
of the system, and 2) a set of representative instantiations for 
the states the system passes through, and the state it settles into, 
after each kinetic Monte Carlo escape.  Executing the model thus 
involves repeating a two-step cycle, first picking a KMC escape time 
for exiting the current state, and then picking a particular instantiation 
of the state-to-state path that the model trajectory follows until 
it settles in some state.  

As will be described in detail below, the first step in the construction 
of the model is to find all MD trajectory segments that spend a certain 
length of time, the {\it dephasing time} ($\tau_{d}$), in a single 
state.  After a trajectory has spent this much time in one state, 
it has lost its memory of how it entered the state and is henceforth 
sampling from a steady-state distribution for that state known as 
the quasi-stationary distribution.  This distribution has the characteristics 
that escape from it is a first-order process, and that when an escape 
occurs, the boundary hitting point is independent of the boundary 
hitting time.  For each state, following these dephased trajectory 
segments forward in time and measuring the number that escape per 
time thus gives the first-order (Markovian) rate constant for escape 
from the state; this rate is used in the KMC step.  The second step 
in the construction of the model involves following each of these 
MD trajectories further forward in time to find the state that it 
settles into (spends a dephasing time in).  The states each of these 
trajectories passes through on the way to the state it settles in, 
and the time it spends in each of these transient states, are stored 
as part of the model.  

With this approach, we can take any definition of the states (though 
some choices would be inefficient), and obtain a compact model of 
the state-to-state kinetics whose accuracy is limited only by the 
completeness of the set of trajectories.  If the states give fully 
Markovian dynamics, then the model will be very compact; for $N$ 
states, it will consist of just $N$ escape rates and $N-1$ final-state 
probabilities for each of these.  When the dynamics on the set of 
states is not Markovian (as is almost always the case), then the 
model will be less compact, by roughly the minimum amount that it 
{\it needs} to be less compact.  For each state, instead of $N-1$ 
final-state probabilities, there will be some number of representative 
{\it instance} trajectories giving the state-to-state path the system 
follows on the way to settling into a final state.  We note that, 
if desired, the information stored for each instance trajectory can 
be more detailed; for example, as we will show for the one-dimensional 
model systems, we can model the time dependence of the trajectory 
position with arbitrary accuracy.  Similarly, for a biomolecular 
system, one could choose to track the evolution through the microstates 
in addition to the macrostates.  (We note, however, that it is typically 
not desirable to use microstates as the main state definitions, because 
the dephasing time to reach the QSD could be extremely long.) For 
systems with very infrequent events, such a model will still be compact, 
as most of the time is spent in the QSD, which can be represented 
with little storage.  

\subsection{Kinetic Monte Carlo (KMC)} \label{kmc}

A molecular system with rare-event dynamics is characterized by occasional 
jumps from one metastable state to another.  In MD simulation of 
materials, it is often the case that events are rare enough so that the 
escape times out of a state are exponentially distributed; the transition probability of escaping a state 
$i$ does not depend on the history prior to entering the state, resulting in a process that is Markovian to a very good approximation.  
For such systems, one can circumvent the time-scale problem in MD 
by devising a stochastic procedure that evolves the system from state 
to state, since the transition depends only on pairwise transition rates  
$k_{ij}$ for moving from state $i$ to state $j$.  If the KMC model 
is parameterized with accurate rate constants (e.g., from transition-state-theory 
calculations~\cite{voterprb}), the state-to-state trajectory thus 
generated accurately mimics an MD trajectory projected onto the metastable 
states.  An important property of a Markov chain is that it gives 
rise to a first-order process that decays exponentially; i.e. the 
probability that the system has not yet left state $i$ is given by 
\begin{equation}
P(t) = exp(-k^{\rightarrow}_{i} t),
\end{equation} 
where $k^{\rightarrow}_{i}=\sum_{l} k_{il}$
is the total escape rate out of state $i$. The probability distribution 
function of the first escape time is given by 
\begin{equation}
\label{expon}
p_{i}(t)=k^{\rightarrow}_{i} \exp(-k^{\rightarrow}_{i} t) .
\end{equation}
Moreover, the exponential distribution holds for each pathway, 
\begin{equation}
p_{ij}(t)=k_{ij} \exp(-k_{ij} t).
\end{equation}
We lay out  the details of the KMC algorithm in Section~\ref{technical}.

\subsection{Quasi-stationary distribution} \label{QSD}

As stated above, most biological systems do not behave in a simple 
Markovian manner.  A trajectory may leave a state shortly after entering 
it, in a direction that is dependent on the way it entered the state.  
In principle, state definitions could be designed to minimize this 
non-Markovian behavior, but there is no guarantee that perfectly 
Markovian states can be constructed.  The quasi-stationary distribution 
(QSD) concept is the key to solving this problem.  The QSD converts 
the escape behavior, for any state definition, to one that is Markovian, 
so that a KMC-like procedure can be employed.  The importance of 
the QSD concept for rare-event dynamics was developed by Le Bris
et al~\cite{qsd1} in a mathematical analysis of parallel-replica 
dynamics (ParRep), a method for parallelizing time in molecular dynamics~\cite{parrep1, 
parrep2}. Recently, the QSD-based approach to ParRep has been applied successfully
to a biochemical system, attaining a parallel speedup of more than 100-fold compared to
direct MD~\cite{lelievrebio}.  The QSD also plays a key role 
in the parallel trajectory splicing method, which we discuss below.  
Here we lay out the mathematics and concepts related to the QSD.  
For complete mathematical proofs and details, readers may refer to~\cite{qsd1}-~\cite{qsd4}.

We consider the scenario of overdamped Langevin dynamics evolving 
on a continuous potential. 
In general, the arguments presented here should also
apply to Langevin dynamics and many other types of dynamics~\cite{qsd1, lelievrebio}
%[{\bf reference here to qsd1 and to Tony Lelievre, private communication}].
The equation of motion is given by: 
\begin{equation}
dX_{t}=-\nabla(X_{t}) dt + \sqrt(2\beta^{-1}) dW_{t},
\end{equation}
where the friction coefficients and the masses have been set to unity, 
$X_{t}$ is a point in configurational space and $dW_{t}$ is the stationary Gaussian process with zero 
mean and unit variance. The phase space is divided into $N$ discrete 
states and it is assumed that the union of all the states covers 
the entire phase space.  Consider a trajectory initiated at $t=0$ 
at a point $X_{0}$ located in state $\alpha$.  If the trajectory 
has not yet escaped this state at a later time $t$, the final configurations 
have a distribution given by 
\begin{equation}
	P_{\alpha}(X, t) = P(X, t | X_{0}=\alpha),
\end{equation}
where $P_{\alpha}(X, t)$ follows the Smoluchowski equation with absorbing 
boundary conditions:   
\begin{equation}
	\partial_{t} P = LP,
\end{equation}
where
\begin{equation}
L=-\nabla \cdot (\nabla V \cdot ) + \beta^{-1} \nabla^{2}. 
\label{defineL}
\end{equation}
$P_{\alpha}(X, t)$ can further be written in terms of the eigenvalues ($\lambda_{i}^{\alpha} > 0$) 
and eigenfunctions ($u_{i}^{\alpha} (X)$) of the linear positively defined operator $L$ 
\begin{equation}
	P_{\alpha}(X,t) = \sum_{i} c_{i} exp(-\lambda_{i}^{\alpha} t) u_{i}^{\alpha}(X),
\end{equation}
where $c_{i}$ are constants such that $P_{\alpha}(X, 0)=\delta (X_{0})$, i.e.,
\begin{equation}
c_{i}=\int \delta (X_{0}) u_{i}(X) d\mu^{-1}(X),
\end{equation}
where $\mu(x)$ is the Boltzmann distribution. 

At long time, the relative contributions from the 
larger
eigenvalues decay away, leading to the approximation for long times
\begin{equation}
     P_{\alpha}(X, t) \approx c_{1} exp(-\lambda_{1}^{\alpha} t) u_{1}^{\alpha}(X)	,
\end{equation}
provided the spectral gap $\Delta_{\alpha} = \lambda_{2}^{\alpha}-\lambda_{1}^{\alpha} > 0$.
The distribution $P_{\alpha}(x,t)$ continues to decay in amplitude, but it 
takes on a constant shape given by $u_{1}^{\alpha} (X)$.  The suitably normalized eigenfunction  
$u_{1}^{\alpha} (X)$ is hence referred to as the quasi-stationary 
distribution, or QSD.  Thus, given a set of trajectories that are 
initiated in state $\alpha$, the subset of the trajectories that 
stay in state $\alpha$ for a long time before moving into the next state, 
such that they have lost knowledge of
how they got into state $\alpha$, give a distribution of 
configurations that is sampled from the QSD.  It has been shown in~\cite{qsd1} 
that once this QSD has been established, the first-escape dynamics 
is Markovian; i.e., escape times are exponentially distributed and 
the hitting point on the state boundary is independent of the escape 
time.  To estimate the time needed to achieve the QSD to good accuracy, 
we note that the slowest relaxation process corresponds to the decay 
of the 2nd eigenvector, so we expect the dephasing to be well achieved 
by a time $t \gg 1/(\lambda_{2}^{\alpha} - \lambda_{1}^{\alpha})$, 
as discussed in Ref~\cite{qsd1}.  We will refer to a time at which 
the system has come very close to the QSD as the {\it dephasing time}, 
$\tau_d$.  

The dephasing time is in general state-specific, so we will 
refer to the dephasing time for state $i$ as $\tau_d^{(i)}$.
We will refer to the QSD escape rate from state $i$, i.e., the expected 
rate at which trajectories escape from state $i$, conditioned on 
these trajectories having been in that state for at least a time $\tau_d^{(i)}$,  
as $k^{QSD}_i$.

\subsection{Dynamical corrections}

As we discussed just above, a trajectory that has stayed 
in a state for longer than $\tau_d$ will make its subsequent escape 
in a Markovian fashion, but the fate of the trajectory after this escape 
will not in general be a simple transition to the adjacent state 
with loss of memory in that state.  For a complex biological system, 
the trajectory may enter and then exit other states in a correlated 
way and with complex dependencies on its past history.  In this section, 
we discuss how this behavior is properly accounted for in our QSD-KMC 
method. 
\redc{We begin with a brief review of dynamical corrections to transition 
state theory; this represents the deep-state Markovian limit, leading 
to a formally defined rate constant between any two states of the 
system.  We then consider the non-Markovian case, for which the QSD 
offers a way to recover Markovian behavior; dynamical corrections 
can again be applied, this time to the QSD escape rate.  However, 
in the case of dynamical corrections to the QSD rate, the resulting 
state-to-state rate constants are no longer unique; they depend on 
the value of the dephasing time, even for long dephasing times.  
Nonetheless, when the dephasing times, the QSD rates, and the state-to-state 
rates are employed in the QSD-KMC procedure, the resulting state-to-state 
evolution is correct.  Thus, QSD plus dynamical corrections offers 
a way to generalize TST plus dynamical corrections for non-Markovian 
systems, but the kinetic evolution of the system can only be predicted 
when this is done in the context of the QSD-KMC approach.
We finish this section with a discussion of the relationship of QSD-KMC 
to the recently developed parallel trajectory splicing method, and we
also present a bootstrap interpretation of QSD-KMC.}

\subsubsection{Dynamical corrections to transition state theory}

For any system, once state boundaries have been defined, we can define 
the transition state theory (TST) rate of escape from each state. 
The TST rate ($k_{ij}^{TST}$) from state $i$ to state $j$ is the 
equilibrium flux through a dividing surface separating states $i$ and $j$.
More specifically, if one counts the number of forward crossings 
at the dividing surface per unit time and divides this quantity by 
the average number of trajectories that are in state $i$ at any time, 
one obtains the TST rate,  $k_{ij}^{TST}$. The total TST escape rate 
out of a state $i$ is given by 
\begin{equation}
k_{i}^{TST} = \sum_{j=1 (j\not= i)}^{N_{esc}} k_{ij}^{TST} ,
\end{equation} 
where $N_{esc}$ is the number of escape pathways from state $i$.  If there 
is a separation of time scales in the system, i.e., if the time to 
lose memory in some state $i$ is much shorter than the average time 
to escape from state $i$, then there exists a proper first-order 
rate constant, $k^{\rightarrow}_{i}$, for escape from state $i$, 
and the TST rate $k^{TST}_i$ approximates this rate constant.  The 
exact rate constant $k^{\rightarrow}_{i}$ can differ from $k^{TST}_i$ 
because while TST assumes that every outgoing crossing of the state 
boundary represents an overall reactive escape event, in reality, 
some of these outgoing crossings may be correlated with prior or 
subsequent boundary crossings.  These correlated events can be in 
the form of recrossings, in which the trajectory quickly reenters 
state $i$, or they may be ``double jump"-type events in which the 
trajectory, while activated, quickly passes through one or more states 
before it finally settles into some state $j$, losing its memory 
there.  
\redc{The number and nature of these correlated events depends on the
characteristics of the system, the temperature, and the Langevin friction
coefficient (higher friction increases the number of recrossings~\cite{kramers}
and decreases the double jump probability~\cite{fer1, fer2, fer3, fer4}).}

Assuming still that there is a separation of time scales, 
i.e., that the duration of these correlated events, $\tau_{corr}$, 
is much shorter than the average time until the next escape out of 
any state ($1/k^{TST}_i$ or $k^{\rightarrow}_{i}$), it is possible 
to account, in a rigorous way, for the effect of these correlated 
events on the rate constants in the system~\cite{chandler, doll}.  
The classically exact rate constant between {\it any} two states 
of the system, adjacent or not~\cite{doll}, can be written as 
\begin{equation}
k_{i\rightarrow j} = k^{TST}_i f_d(i \rightarrow j) ,
\end{equation}
where $f_d(i \rightarrow j)$ is a final-state-specific dynamical 
correction factor, which can be computed from the results of properly 
sampled half-trajectories initiated at the boundary to state $i$~\cite{doll}.  
Half-trajectories that are in state $j$ at time $t=\tau_{corr}$ contribute 
positively to $f_d(i \rightarrow j)$ if they were exiting state $i$ 
at $t=0$ and contribute negatively if they were entering state $i$ 
at $t=0$. Specifically, 
\begin{equation}
f_d(i \rightarrow j) = (2/N_{traj}) \sum_I^{N_{traj}} \gamma(I) \Theta(I,j,\tau_{corr}) ,
\end{equation}
where $N_{traj}$ is the number of sampled half-trajectories, $\gamma(I)$ 
indicates the phase of half-trajectory $I$ (+1 if exiting state $i$, 
-1 if entering state $i$, at $t=0$), and $\Theta(I,j,t)$ is the Heaviside 
function for presence in state $j$ at time t (Note that in Ref.~\cite{doll}, 
a factor of $1/N_{traj}$ was inadvertently omitted from the RHS of 
the analagous Eq. (4.6)).  Thus, with TST as a starting point, and 
assuming a separation of time scales, this many-state dynamical corrections 
formalism provides a definition for the classically exact rate constants
between all pairs of states in the system.  Once these rates are known, 
a KMC procedure can be used to advance the system from state to state~\cite{voterprb}.  

\subsubsection{Dynamical corrections in QSD-KMC} \label{dyn-corr}

In the QSD-KMC method we are presenting here, we take an approach 
analogous to the dynamical corrections to TST described just above, 
but with a key difference:  we are interested in systems 
for which we cannot assume that we have a clean separation of time scales, 
i.e., systems that may not be Markovian.  For such a system, although 
the TST rate of escape from any state ($k^{TST}_i$) is still 
well defined as the equilibrium escape flux, and could be calculated 
if desired, it is no longer a meaningful escape rate. If the escape 
behavior is not Markovian, there is no simple first-order rate constant.  
This is where the QSD concept comes into play.  For the base escape 
rate from state $i$, instead of using $k^{TST}_i$, we use $k^{QSD}_i$, 
as defined in Section~\ref{QSD}; this gives a good first-order rate 
constant for escape from state $i$, but only after the system has 
spent a time of at least $\tau_d^{(i)}$ in that state.  Then, in 
analogy with the use of half-trajectories to correct TST for the 
effect of correlated events, we follow trajectories escaping from 
the QSD until they settle in some state $j$.  The requirement for 
settling in a state $j$ is that the trajectory spends a time $\tau_d^{(j)}$ 
in that state.  We note that it is possible that the correlated event 
may bring the system back to state $i$, so that $j=i$. While in dynamical 
corrections to TST, this type of return event merely contributes 
to a lowered overall rate of escape, in the QSD-KMC algorithm this 
type of returning trajectory is a meaningful event, which we track 
through the various intermediate states the trajectory visits before 
returning to state $i$.

We will refer to these dynamical correction trajectories for QSD-KMC 
as {\it instance} trajectories, or {\it instances}.  When the system 
escapes from the QSD of some state, it has a probability of settling 
in a certain final state, as well as probabilities for the various 
paths through intermediate states along the way.  The instance trajectories, 
harvested as internal segments of the underlying MD trajectories 
from which the QSD-KMC model is built, give us different possible 
instantiations for this process.  The number of instances that take 
the system from state $i$ to state $j$, divided by the total number 
of instance trajectories for state $i$, gives the probability that 
an escape from state $i$ leads to state $j$.  We note that while 
the QSD rate constants \{${k^{QSD}_i}$\} are independent of the choices 
for the dephasing times (provided they are chosen to be large enough), 
the instance trajectory sets, and their associated rate constants, 
are not.  This is an important point that will be discussed in Section~\ref{rate}.

\subsubsection{Connection to parallel trajectory splicing} \label{parsplice}

We note that this QSD-KMC approach makes a connection to the recently 
developed parallel trajectory splicing (ParSplice) method~\cite{parsplice}.  
ParSplice is a generalization of parallel replica dynamics; both 
methods parallelize time to accelerate molecular dynamics of rare 
events without requiring prior knowledge of the possible events.  
\redc{These methods are usually applied to material systems for which each 
potential basin of attraction defines a state, and the total number
of states the system may visit as it proceeds is typically huge.} 
In ParSplice, multiple trajectory segments are simultaneously propagated 
in one or more known states of the system; these segments are harvested 
and spliced end to end to make a trajectory that reaches long times.   
To be a valid, spliceable segment of time length $t_{segment}$, the 
trajectory must have remained in a single state for a time $\tau_d$ 
prior to the beginning of the segment, and also must remain in a 
single state (the same or another) for the final $\tau_d$ of the 
segment.  Spliceable segments thus fall into two classes: those that 
remain in one state the whole time (very common for a rare-event 
system) and those that make one or more transitions during the segment.  
We see then that transitioning segments in ParSplice are exactly 
the same as the instance trajectories defined for QSD-KMC.  
\\
%In the ParSplice approach, the general definition of a state is set before 
%the simulation begins (typically it is defined as the basin of attraction 
%of a local minimum in the potential, which works well for a material 
%system), and careful application of the ParSplice procedure in a 
%parallel computing environment gives MD-accurate evolution of the 
%system from state to state~\cite{parsplice}.  
%%%%%%%%No prior knowledge is required about where the ParSplice 
%%%%%%%%trajectory may lead or what states may be visited.  
%
%In contrast, in the QSD-KMC approach, we are post-analyzing a set 
%of MD trajectories, typically for a biological system in which a 
%pre-set state definition such as the local basin of attraction would 
%be inappropriate.  In the post-analysis we have the freedom to define 
%the state space in any way we want, although this choice may impact the
%short-trajectory generation procedure, as discussed in Section~\ref{corr-instance}.
%Given this set of states, and 
%the trajectory information transformed into QSD rates and correlated-event 
%instances, the QSD-KMC procedure advances the system stochastically 
%from one state to the next.  In general, the state-to-state 
%evolution will be very accurate (i.e., probabilistically matching 
%what a long MD trajectory would give) for time scales that remain 
%shorter than the total time accumulated in the set of MD trajectories.

\subsubsection{Bootstrap Interpretation}

Any trajectory can be segmented into excursions each time a dephasing 
event occurs. The loss of memory of the past after a dephasing event implies that the
excursions are independent and identically distributed conditionally on the state in which 
the dephasing event occurred. Given an ensemble of excursions, we can construct 
a bootstrap path by resampling the excursion from the appropriate subset (given the initial state).
This interpretation shows that the statistical accuracy will depend on the number of excursions 
starting in each state.  

\section{Algorithm and Practical Aspects of Implementation} \label{technical}

%{\bf (revisit this paragraph when we are all done, to make sure it correctly says what we cover)}
\redc{In this section, we describe the practical aspects of implementing 
the QSD-KMC method.  We first discuss the Anderson-Darling test
procedure that we use for estimating the dephasing time. Next, we discuss how 
the pairwise rate constants are calculated in this model. 
We then lay out the procedure for generating the short
trajectories for constructing the QSD-KMC model. 
Finally, we discuss the QSD-KMC algorithm in detail and how it can be applied in practice 
to generate exact state-to-state evolution in a biological system. }

\subsection{Choice of dephasing time} \label{choice}

Choosing an appropriate dephasing time is a key part of the QSD-KMC 
algorithm. 
%While the dynamics can be made arbitrarily accurate in 
%the limit of infinite dephasing time, for practical applicability, 
%there needs to be a tradeoff between accuracy and statistics available 
%from the underlying MD simulations; a longer dephasing time uses 
%up a larger fraction of the total length of the MD trajectories, 
%leaving fewer segments from which the state-to-state behavior can 
%be represented.  
For very-low-dimensional systems, it is possible 
to directly calculate a good estimate of the dephasing time by appealing 
to the expression for the error in the QSD derived by Le Bris et 
al.~\cite{qsd1},  $\Delta\mu_1^{\alpha} \approx \exp(-(\lambda_{2}^{\alpha}-\lambda_{1}^{\alpha})\tau_{d})$, 
%{\bf (I changed this equation in two ways - please verify)}
where $\lambda_1^{\alpha}$ and  $\lambda_2^{\alpha}$  are the first two eigenvalues 
of the Fokker-Planck equation with absorbing boundary conditions, 
as discussed above.  
For higher-dimensional systems, where it becomes 
unfeasible to solve the Fokker-Planck equation, direct calculation 
of $\tau_d$ is extremely difficult. 
%%%Although some approximate approaches 
%%%have been proposed~\cite{qsd2}, in general this is well known as 
%%%a difficult problem, which we will not try to make any further advance 
%%%on here.  
For the purpose of demonstrating the QSD-KMC method, we estimate 
$\tau_d$ for each state as the point in time when the survival probability 
becomes exponential. 
We note that, as pointed out by Le Bris et al~\cite{qsd1}, this does 
not guarantee the system has achieved the QSD.  However, we believe 
this approach will suffice for our purposes here. Indeed, in the 
results section below, we will find that using this approach we obtain 
very accurate results. The QSD-KMC trajectory predictions agree extremely 
well with the results calculated from the underlying MD trajectories.  
For systems with very deep states, it is no longer feasible to use
the onset of exponentiality to define the dephasing time, as it requires
running each trajectory long enough to see an escape.  In this situation,
we suggest using the Gelman-Rubin approach~\cite{gelman}
{as proposed by Binder et al~\cite{qsd2}
and recently implemented by Hedin and Lelievre~\cite{lelievrebio}.

\begin{figure}
\centering     %%% not \center
\subfigure[$\tau_{d}^{M}=81.0$ $ns$, $k_{M}^{QSD}=0.0085$ $ns^{-1}$]{\label{survival-vil-1:a}\includegraphics[width=60mm]{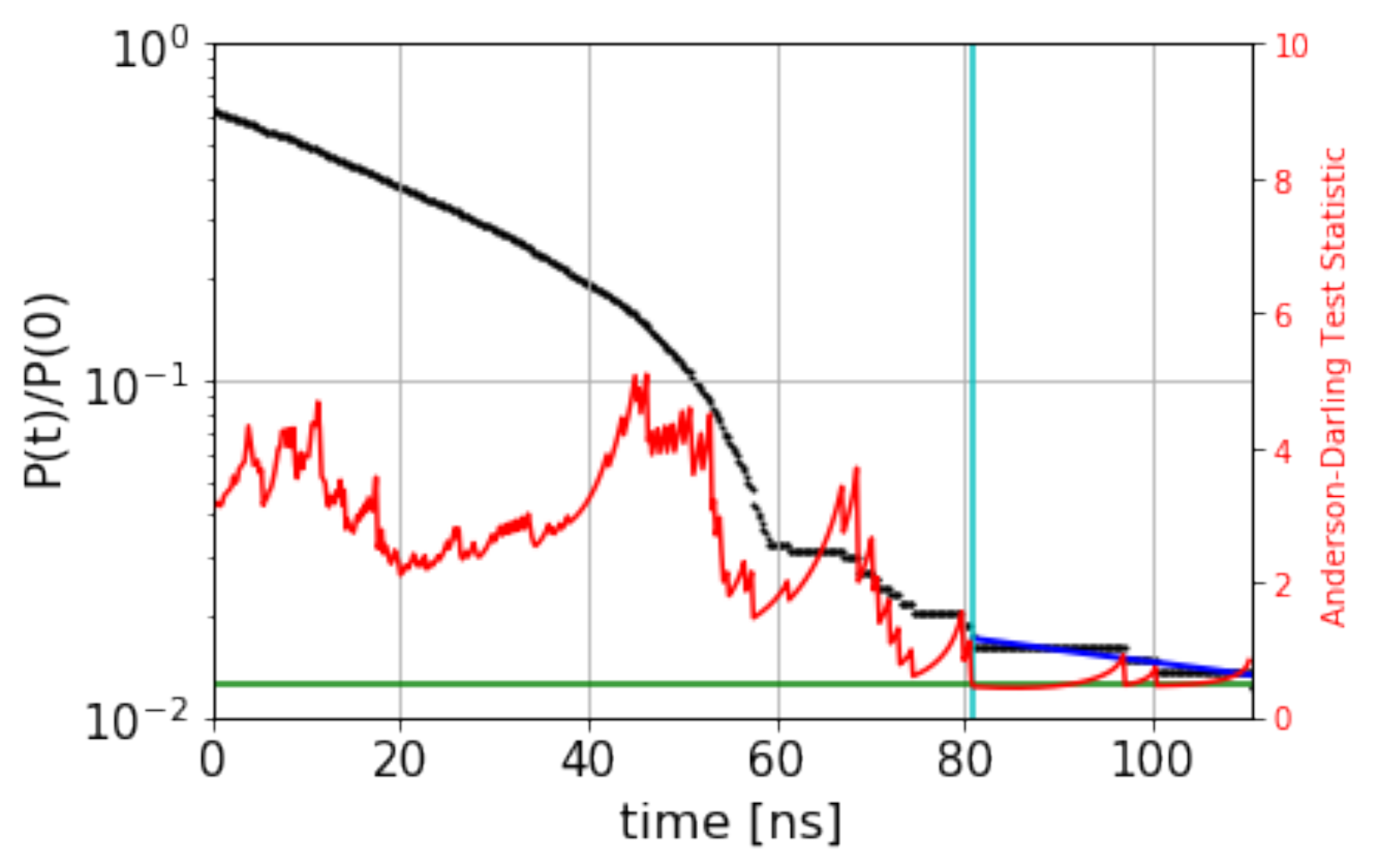}}
\subfigure[$\tau_{d}^{F}=62.6$ $ns$, $k_{F}^{QSD}=0.0095$ $ns^{-1}$]{\label{survival-vil-1:b}\includegraphics[width=60mm]{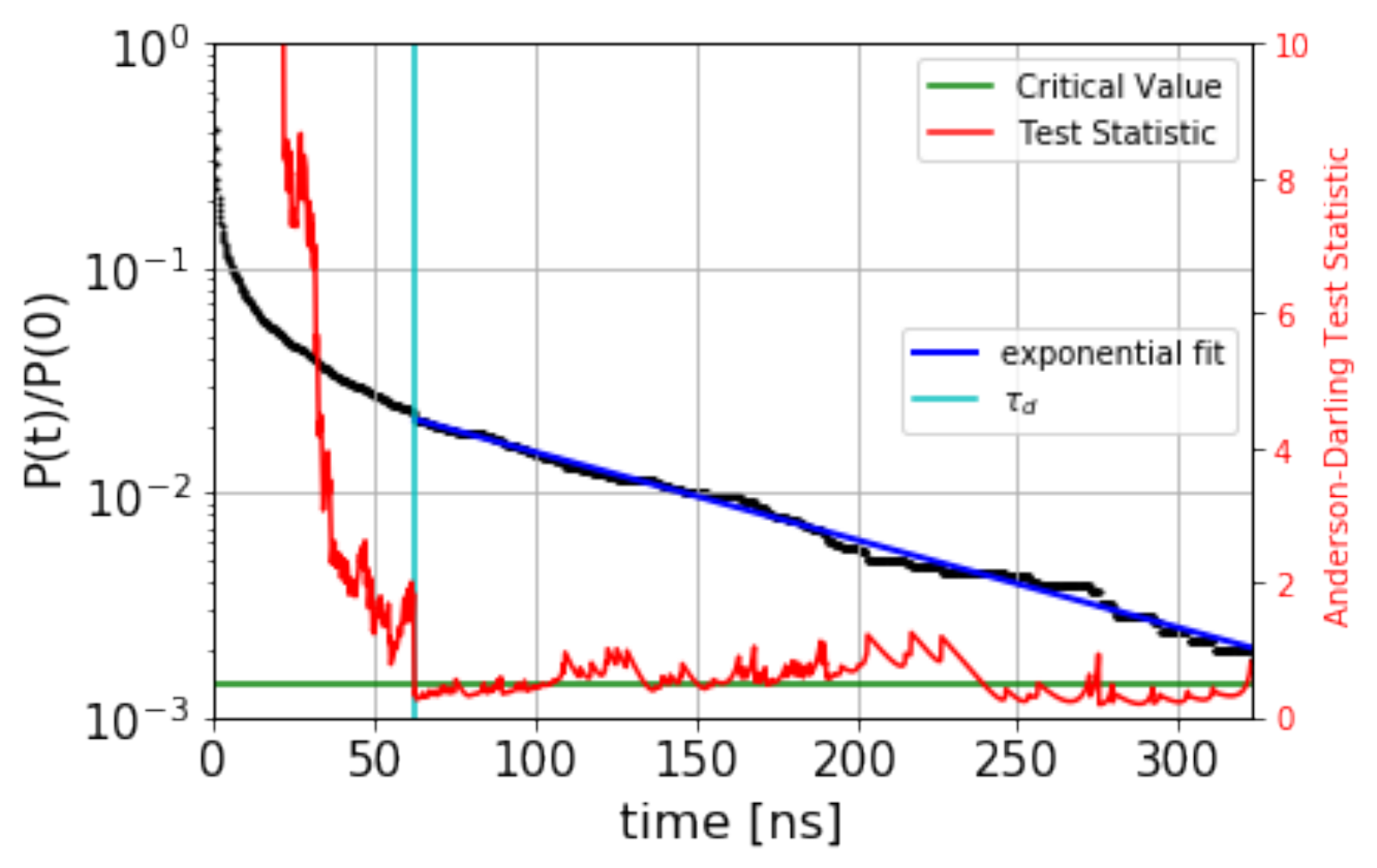}}
\subfigure[$\tau_{d}^{U}=30.4$ $ns$, $k_{U}^{QSD}=0.0052$ $ns^{-1}$]{\label{survival-vil-1:c}\includegraphics[width=60mm]{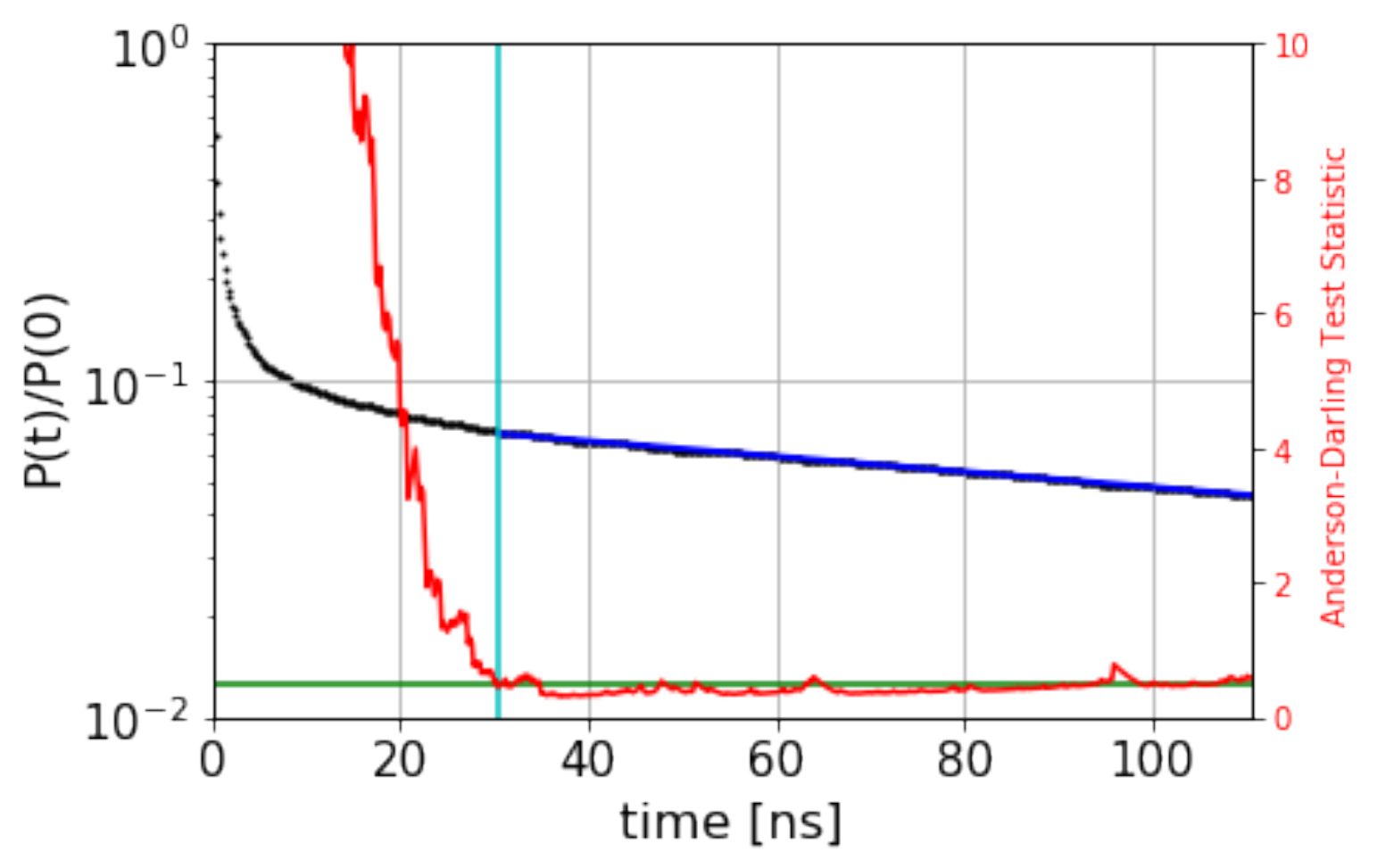}}
\caption{Survival probability function (probability that the system 
has not yet escaped from state $i$), and Anderson-Darling determination 
of the dephasing time, for three PCCA states of villin headpiece.  The 
survival probability function is indicated by black dots, the Anderson-Darling 
test statistic is in red (right-hand scale), the Anderson-Darling 
critical value is shown in green (right-hand scale), and the chosen 
dephasing time is the vertical cyan line.  An exponential fit to 
the survival probability for times greater than the dephasing time 
is shown as a blue line.  The dephasing time is set to the shortest
time for which the Anderson-Darling test statistic is below the critical 
value of 0.5; for longer times, the survival probability is likely exponential,
indicating the QSD has been reached to a good approximation. } 
\label{survival-vil-1}
\end{figure}

We automate the determination of the onset of exponentiality by employing 
the Anderson-Darling test~\cite{anderson} for an exponential distribution.  
Normally, the Anderson-Darling test is used to rule out, to high 
confidence, that data comes from a given distribution, by comparing 
a test statistic, computed from the data, to a critical value $\alpha$ that 
depends on the desired confidence; if the test statistic 
is greater than the critical value, then the null hypothesis (that the 
data is exponential) is rejected. 
(For example, $\alpha$=$\sim$1.3 for 95\% confidence, or a $p$ value of 0.05.)  
Here we use the Anderson-Darling 
test in the opposite sense, seeking reassurance that the data is not too 
likely to violate exponentiality.  We achieve this by requiring 
that the standard test statistic~\cite{anderson} be below a much lower critical value 
of $\alpha=0.5$. We determined this critical value empirically:
data drawn randomly from a true exponential gives a test statistic that
is below $\alpha$=0.5 roughly 50\% of the time. For computing the test
statistic, we use the standard scipy.stats library in python. 

The automated procedure then works as follows.
We scan candidate values of the dephasing time, $\tau_d^c$, computing
the Anderson-Darling test statistic for the escape-time data with 
escape times less than $\tau_d^c$ discarded. The 
dephasing time $\tau_d$ is then set to the lowest value of $\tau_d^c$ 
for which the test statistic is less than 0.5.
As an example, Figure~\ref{survival-vil-1} shows the survival probability 
function and the Anderson-Darling-determined dephasing times for 
the three states of villin headpiece.
(These states were obtained via the application of MSM and Perron 
Cluster-Cluster Analysis (PCCA)~\cite{pccaweber}; 
\redc{the free energy map of villin headpiece and the three states are
shown in the figures in Section~\ref{discussion}.}
More details on how we characterize these states as misfolded, folded and unfolded
are also given in Section~\ref{discussion}.) 
We observe that the Anderson-Darling test procedure is quite sensitive to the nonexponentiality, and appears to give a fairly conservative
determination of $\tau_d$, a desirable characteristic.
%%%It can be seen that the survival probability function and the Anderson-Darling 
%%%test statistic plot for State $C_{7}^{ax}$ that a dephasing time 
%%%of 162 $ps$ appears to be a valid choice. 

\subsection{Calculation of rate constants}
\label{rate}
Once the dephasing time for a state has been estimated, the first-order 
escape rate out of the QSD for that state, $k_{i}^{QSD}$, is given 
by the negative slope of the log of the survival probability for 
times greater than $\tau_d$.  Alternatively, one can calculate this 
rate as the average escape rate for the trajectories after time $\tau_d^{(i)}$,
\redc{provided all these escape times are known (i.e., that no trajectories are prematurely terminated).}
By definition, these rates do not change (or should change very little) 
as the dephasing time is further increased.  The QSD escape rates \{$k_{i}^{QSD}$\},
along with the instance trajectories, are all that is needed to carry out QSD-KMC
simulations. However, before leaving this section, we make a few other points
about rates in this type of model.   

Once first-order QSD escape rates out of a state have been estimated, 
the pairwise rate constants accounting for the correlated dynamical 
events can be defined as
\begin{equation}
\label{rates}
%%%%k^{settle}_{ij} = k_{i}^{QSD} \times p^{settle}_{ij} = k_{i}^{QSD} \times \frac{\text{No.\,of\,instances\,that\,settle\,in\,state\, `j' \,starting\,in\,state\, `i'}}{\text{No.\,of\,instances\,that\,settle\,in\,any\,state\,starting\,in\,state\,`i'}},
k^{settle}_{ij} = k_{i}^{QSD} \times p^{settle}_{ij} = k_{i}^{QSD} \times \frac{\text{No.\,of\,instances\,that\,settle\,in\,state\,} j\  \text{starting\,in\,state\, }i}{\text{No.\,of\,instances\,that\,settle\,in\,any\,state\,starting\,in\,state\,}i},
\end{equation}
where $p^{settle}_{ij}$ is the probability that, starting in state 
$i$, the trajectory first settles (resides for at least a dephasing 
time, as discussed above) in state $j$.  We note that $p^{settle}_{ij}$ 
is analogous to $f_d(i \rightarrow j)$, discussed above for the case 
of dynamical corrections to TST in a system with a clear separation 
of time scales.  However, there is an important difference: 
$p^{settle}_{ij}$ (and hence $k^{settle}_{ij}$) depends on the choice 
of dephasing times for the states, even when all the dephasing times 
are appropriately large.  That $k^{settle}_{ij}$ will vary with the 
dephasing time can be easily understood from a simple thought experiment 
on a three-state system with states 1, 2, and 3.  Starting with a 
conservative (long) value for the dephasing time for all the states, the 
rate constants can be calculated using Eq.~\ref{rates}. Focusing 
on $k^{settle}_{12}$, if we further increase $\tau_d^{(2)}$, 
the probability that the system settles in state 2 will be 
reduced; it becomes more likely that the system fails to stay long enough 
in state 2 to be declared as settled, and instead 
makes additional transitions, ultimately settling with greater likelihood 
in state 1 or 3.  Thus, $k^{settle}_{12}$ is reduced as we increase $\tau_d^{(2)}$. 

We can also define pairwise rate constants ($k^{hit}_{ij}$) corresponding 
to the state first encountered upon escape from the QSD,
\begin{equation}
k^{hit}_{ij} = k_{i}^{QSD} \times p^{hit}_{ij} = k_{i}^{QSD} \times \frac{\text{No.\,of\,instances\,that\,initially\,hit\,state\,}j \text{ as\,they\,exit\,state\,}i}{\text{No.\,of\,instances\,that\,escape\,state\,}i}.
\end{equation}
These rates are {\it not} dependent on the dephasing times (once the dephasing 
times are large enough to give accurate QSDs), since the hitting 
point on the state boundary is independent of the escape time once 
the QSD is established~\cite{qsd1}.   Thus, although $k_{i}^{QSD}$ and \{$k^{hit}_{ij}$\} 
are fundamental properties of the system once the state boundaries 
are defined, the settling rates \{$k^{settle}_{ij}$\} are not. 

%To illustrate this, Table~\ref{table-1} and Table~\ref{table-2}  show selected rates for the alanine dipeptide system.
%Table~\ref{table-1} shows the total escape rate out of state $\alpha_{R}$, $k_{\alpha_{R}}^{QSD}$, 
%and the pairwise hitting rate constants $\left\{ k^{hit}_{\alpha_{R} (j)}\right\}$ 
%as a function of $\tau_d^{\alpha_{R}}$, while keeping the dephasing times 
%for the other states fixed at their Anderson-Darling values. 
%We note that even though $\tau_{d}^{\alpha_R}$ = 664 $ps$, 
%$k_{i}^{QSD}$ and the $\left\{k^{hit}_{\alpha_R (j)}\right\}$ values essentially do not change
%even for shorter dephasing times. 
%%for dephasing times greater than 100 $ps$.
%This is consistent with the conservative value for $\tau_{d}$ obtained from the Anderson-Darling procedure.
%In Table~\ref{table-2}, we show the pairwise rate constants $\left\{k^{settle}_{\alpha_{R} 
%(j)}\right\}$ as a function of dephasing time of the final state, $\tau_d^{(j)}$, while
%keeping the dephasing times of all other states at their Anderson-Darling value.  
%It can be clearly seen that these 
%rates decrease as $\tau_d^{(j)}$ increases.  It can also be seen 
%that for the shortest possible dephasing time (i.e. $\tau_{d}$=2 
%$ps$) for the final state, $\lim_{\tau_{d}^{j} \to 0}$ $k^{settle}_{\alpha_{R} 
%(j)}$ $ > $ $k^{hit}_{\alpha_{R} (j)}$. This is expected, because 
%any trajectories escaping from state $\alpha_{R}$ that first hit a state different than $j$, 
%but do not settle there, have a chance 
%of settling in $j$. 

To summarize, the QSD escape rates \{$k_{i}^{QSD}$\} generated for 
use in the QSD-KMC model are unique once the state boundaries have 
been specified and suitably long dephasing times have been determined.  
Adjacent-state hitting probabilities (not needed for the QSD-KMC procedure) are also unique.
However, for a complex biological system, the settling rates \{$k_{ij}^{settle}$\} 
will in general have no unique value; they will continue to change 
as the dephasing times are increased.  This is a natural consequence 
of the non-Markovian nature of these systems.  The key point, though, 
is that in spite of this characteristic, using these settling rates 
in the QSD-KMC procedure will generate appropriate and accurate predictions 
of the state-to-state dynamics, for any (safe) choice of dephasing times.

\redc{
\subsection{Generating and Analyzing the Short Trajectories} \label{corr-instance}

Here we specify the procedure for generating the short trajectories 
such that they are long enough (and just long enough) to allow extraction 
of the information for constructing the QSD-KMC model.  We assume 
that the user has chosen a set of reaction coordinates and defined the states
in that subspace. 
There are two stages in the method: 

%{\bf Animesh -- I have make it into a pseudo-code list of steps. 
%Would you mind putting it back into the list form for me? 
%I am realizing I don't know how to do that, and for some reason I cannot
%find a copy of the version in which you had it that way.  If you 
%prefer, I can turn it into sentences-- I am fine with either way, but the
%list might be clearer or crisper.  Thanks.} 
\textbf{Stage 1: Determining the dephasing times and QSD escape rates.}

For each state $i$:
\begin{itemize}
\item Initiate a number of trajectories.
\item Integrate each of these trajectories forward in time until it escapes from state $i$.
\item Using this list of escape times, apply the Anderson-Darling procedure discussed above
to determine the appropriate dephasing time $\tau_{d}^{(i)}$.  Also determine 
the QSD escape rate $k_{i}^{QSD}$ for this state.
\end{itemize}

\textbf{Stage 2: Creating the instance trajectories.}

For each state $i$: 
\begin{itemize}
\item Discard the trajectories initiated in state $i$ that escaped before $\tau_d^{(i)}$. 
\item For each of the remaining trajectories,
continue integrating the trajectory forward in time until it 
dephases (settles) in some state $j$, where $j$ may be the same as $i$.  
Record the desired information about the instance trajectory, from the 
time it left state $i$ until the time it finished dephasing in state $j$.
\end{itemize}
We note that this procedure generates a set of trajectories with varying 
lengths.  It is generally not sufficient to generate a set of short 
trajectories that are integrated for a fixed time unless the fixed time 
is fairly long.  If any trajectories 
are truncated before they escape from their initial state, it becomes 
difficult to properly determine the QSD escape time and the dephasing 
time.  Moreover, if any trajectories are truncated before settling 
into a final state, a bias will be introduced into the QSD-KMC model, 
because instance trajectories that are long (i.e., that visit many 
intermediate states) are then less likely to be included in the model.}

As mentioned above, for systems with very deep states, it may not be
feasible to use the onset of exponentiality to define the dephasing 
time, as it requires running every trajectory long enough to see an 
escape.  In this situation, the procedure could be modified as follows:

\textbf{Stage $1^{\prime}$: Determining the dephasing times and escape times.}

For each state $i$:
\begin{itemize}
\item Initiate a number of trajectories.
\item Integrate these trajectories forward in time, replacing any that escape from state
$i$ using a Fleming-Viot cloning procedure~\cite{fleming1, fleming2}. 
%{\bf - xxx put in refs 46 and 47 from the Hedin and Lelievre, citations for Fleming-Viot)}
\item Determine when dephasing has been achieved using the Gelman-Rubin criterion (this also defines $\tau_{d}^{(i)}$).
\item Continue these dephased trajectories forward in time
until some number of them have escaped. The QSD escape time ($1/k_{i}^{QSD}$) 
can be determined by averaging these escape times, noting that each 
successive escape time, relative to the previous escape time, should 
be reduced by a factor of the number of dephased trajectories remaining 
in the state, to account for the parallelization speedup 
(as in parallel-replica dynamics~\cite{parrep1}).  
\end{itemize}

\textbf{Stage $2^{\prime}$: Creating the instance trajectories.}

For each state $i$: 
\begin{itemize}
\item For each of the trajectories that escaped after dephasing, 
continue integrating the trajectory forward in time until it 
dephases (settles) in some state $j$, where $j$ may be the same as $i$.  
Record the desired information about the instance trajectory, from the 
time it left state $i$ until the time it finished dephasing in state $j$.
\end{itemize}

\subsection{Algorithm}

\textbf{KMC algorithm:} Suppose the system is in state $i$; there 
are $N$ pathways along which the system can make a transition. The 
probability distribution of the first escape time for each of the 
pathways is given by $p_{ij}=k_{ij} \exp(-k_{ij} t)$. One can draw 
a time $t_{ij}$ from the exponential function for each of the pathways 
via  $t_{ij}=-(1/k_{ij}) \ln(r)$ where $r$ is a real random number between 
0 and 1. This time is representative of a first escape time for a 
first-order process. One then finds the pathway $j_{min}$ that has 
the minimum value of $t_{ij}$, the system is moved to state $j_{min}$, 
and the simulation clock is advanced by $t_{ij}$. The more efficient 
version of this algorithm involves stacking all the rate-constant 
objects (where each object has a length equal to $k_{ij}$) end to 
end and choosing a random number between 0 and $k^{\rightarrow}_{i}$ 
(where $k^{\rightarrow}_{i}=\sum k_{ij}$ ). The object associated 
with the chosen random number is the pathway that the system follows 
and the simulation clock is advanced by time $t=1/(k^{\rightarrow}_{i}) 
\ln (r)$. This approach is referred to as the Bortz, Kalos and Lebowitz 
(BKL) algorithm~\cite{bkl}.  

\textbf{QSD-KMC algorithm:} The idea is quite similar to the BKL algorithm 
described above.  Suppose the system is in state $i$. We draw an 
exponentially distributed time using the total escape rate $k_{i}^{QSD}$: 
$t = -(1/k_{i}^{QSD}) \ln(r)$, and advance the simulation 
clock by this time. Next, we pick a random position along an array 
of trajectory instances for state $i$.  The chosen instance defines 
the pathway that the system follows through intermediate states and 
the state that it finally settles in.  The simulation clock is advanced 
by the total time the instance trajectory spends going through all 
the intermediate states as well as the dephasing time of the final 
state $j$ into which it settles. The system is then moved to the 
new state $j$ and the process is repeated again with the list of 
pathways and rates for state $j$. A schematic representation of the QSD-KMC procedure
is shown in Figures~\ref{cartoon:a} and ~\ref{cartoon:b}.
\begin{figure}
\centering     %%% not \center
\subfigure[Procedure for extracting instances from short MD trajectories]{\label{cartoon:a}\includegraphics[width=95mm]{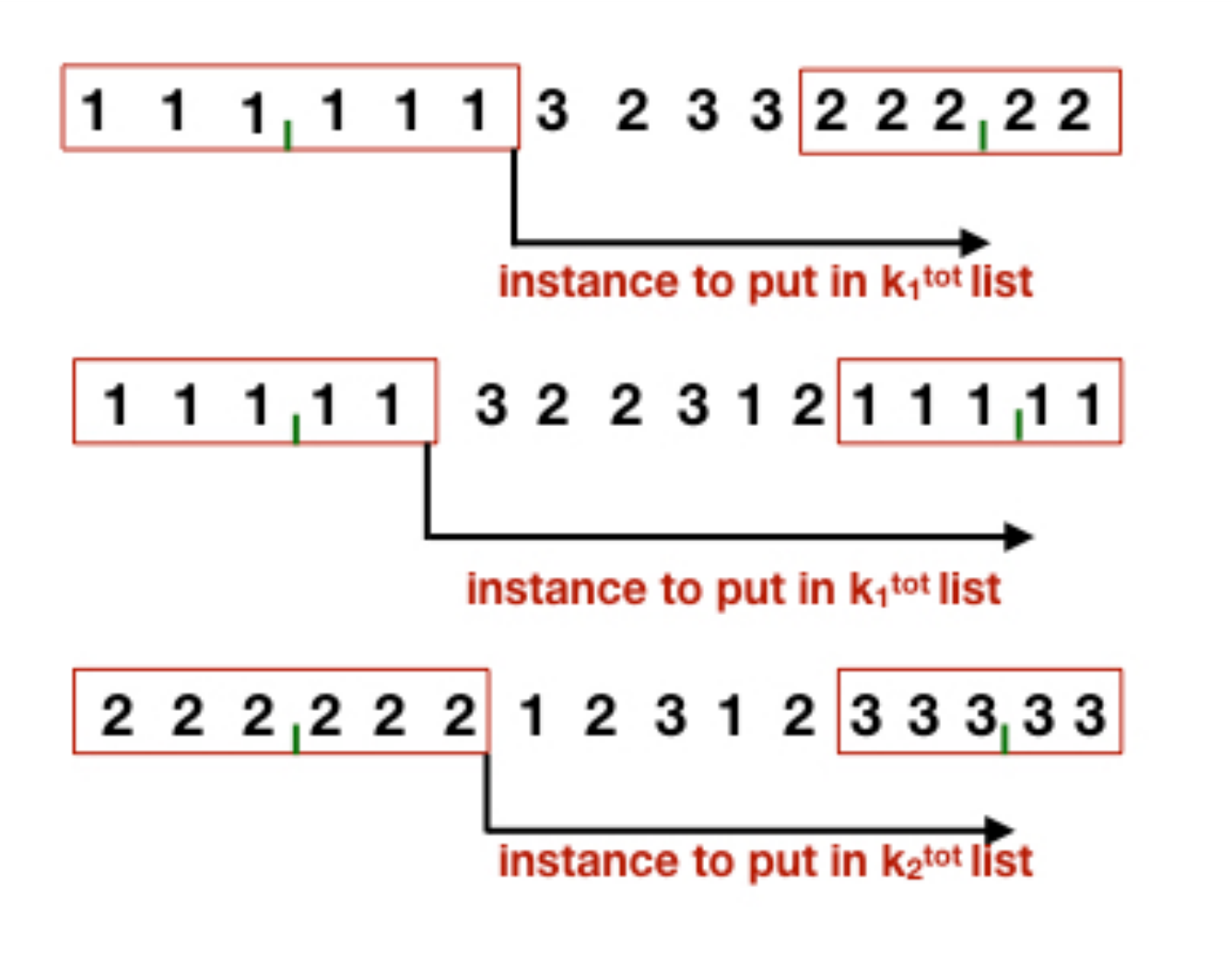}}
\subfigure[QSD-KMC algorithm]{\label{cartoon:b}\includegraphics[width=155mm]{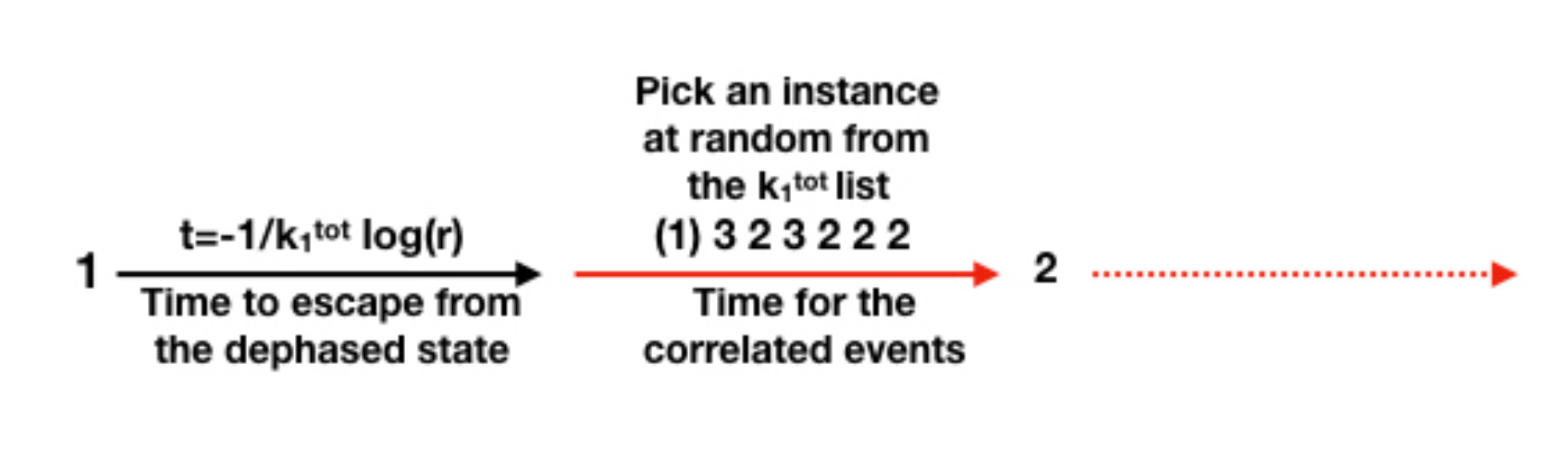}}
\caption{Schematic illustration of algorithmic procedures.  We consider 
three states (1, 2 and 3) and we choose the 
dephasing time for these states as 3 time units. \redc{a) Procedure for extracting
instances from short MD trajectories, once they have been projected onto these states.}
b) QSD-KMC procedure.
}
\label{cartoon}
\end{figure}

QSD-KMC predicts the state-to-state evolution with accuracy that improves 
with increasing $\tau_{d}$ and total underlying trajectory time $T$. Similar to 
the other state-of-the art methods such as MSM, the long time dynamics      % xxxx
is predicted from a compact model built from a database of short 
MD simulations, although there is an additional constraint regarding the length 
of the individual MD trajectories that depends on the macrostates, as discussed above
in Section~\ref{corr-instance}.
%
%%Ideally, a trajectory should be long enough that at least twice during 
%%its journey it spends more than a dephasing time in some state 
%%(the same state or two different states).  Such a trajectory will 
%%contribute both to estimation of one or more QSD escape rates, and 
%%will provide one or more instance segments.  Trajectories that spend 
%%a dephasing time in just one state can only contribute to the estimation 
%%of the QSD escape rate from that state.  Trajectories that never dephase
%%in any state are discarded. \redc{We note here that if the trajectories which
%%initially settle into some state but later do not dephase into any state are 
%%discarded, then it will induce a bias in the model. In view of 
%%this issue, we propose a procedure for generating
%%short MD trajectories for computing long-time dynamics
%%using QSD-KMC:

There are two main sources for errors in QSD-KMC. The first one is the statistical 
error arising from observing only a finite number of transitions 
from one dephased state to another, which leads to uncertanities 
in the estimated rate constants.  One way to reduce this error is 
to leverage an adaptive-sampling approach~\cite{adapt}, wherein by 
observing the uncertanities in the rate constants, one can initiate 
additional simulations in states that have a large contribution to 
the uncertainity.  The other source of error is the increasing uncertainty 
in the QSD-KMC predictions at longer times, due to the possibility 
of missing rates and pathways~\cite{abhi1, abhi2} on these timescales.  
Following Chatterjee and coworkers~\cite{abhi1, abhi2}, we can define 
the validity time of a KMC model, i.e. the time duration for which 
the KMC model continues to give correct dynamics compared to the 
underlying MD dynamics, after which the predictions are less reliable.  
This error can be reduced by employing some of the techniques described 
in~\cite{abhi3} to boost the validity time of the model.

\section{State Optimization} \label{optimization}

Here we propose a Metropolis Monte-Carlo method that optimizes the 
shapes of the macrostates, such that the states obtained correspond, 
in some sense, to the set of states that are maximally metastable.  
The procedure is based on the minimization of the total time of the 
correlated events in a long QSD-KMC trajectory normalized by the 
length of the trajectory.  Consistent with the discussion above, 
the correlated event time begins upon escape from a state in which 
the trajectory spent at least a dephasing time, and ends when the 
trajectory has spent at least a dephasing time in some (perhaps the 
same) state.  For a given number of macrostates ($M$), the procedure 
thus identifies the partitioning of the state space into $M$ contiguous 
states that minimizes the total correlated event time in a given 
set of MD trajectories.  

\redc{While the procedure we will describe can in principle be applied to any 
system studied with QSD-KMC, in fact it may not be so straightforward 
when one has pre-generated a set of short trajectories.
This is because as the state definitions change during the optimization, 
strictly obeying the two-stage procedure described in Section~\ref{corr-instance} 
dictates re-evaluating the dephasing time for each state, and this 
in turn may require extending the length of some trajectories to 
re-satisfy the requirement that every trajectory fully settles in 
a final state. The optimization procedure is thus coupled to the running
of the trajectories.  On the other hand, there may be effective ways to
achieve the optimization approximately with minimal modification of the
trajectories.  Also, sometimes one has available very long trajectory
segments, as is the case for the alanine dipeptide example below, so that
multiple instance segments can be taken from a single trajectory, and each
instance segment can be extended as necessary with only a very low probability
of coming to the end of the trajectory.  In this case only a very 
small bias error is introduced by the finite trajectory length.  Finally, for the
case in which a very long, single trajectory is available, as is the case
for the villin headpiece example below, the procedure can be applied 
with essentially no error arising from the finite trajectory length.}

The process is initiated by providing a starting guess for the definition 
of the states, perhaps from PCCA, or from a simple geometric 
partitioning, in terms of the basis of the $n$ microstates.
Given this initial lumping of microstates 
into macrostates, we calculate the average fraction of the QSD-KMC 
trajectory time that is involved in correlated events.  We will refer 
to this as the fractional ``outside" time, $f^{out}$, because the 
trajectory is outside of any QSD for this fraction of the time.  
If the MD trajectory information underpinning the QSD-KMC model is 
``balanced," i.e., if the trajectories are each long enough to 
have visited all the states many times, or if the number of trajectories 
initiated in each state is proportional to the equilibrium population 
of that state, then $f^{out}$ can be computed directly 
from the state-projected MD trajectory information, without generating 
any QSD-KMC trajectories.  In this case, $f^{out}$ is then simply given by the total instance 
time divided by the total MD trajectory time that was usable in generating 
the instance information.  We exploit this approach in the present work. 
If the trajectory information is not balanced, then $f^{out}$ 
can be computed from a sample set of QSD-KMC trajectories or, alternatively, 
it can be computed for each state $i$ from 
$k^{QSD}_i$ and the average length of the instance trajectories escaping 
from state $i$, and these can be weighted by the equilibrium population 
fractions obtained from the first eigenvector computed from diagonalization 
of the appropriate transition matrix.  

To generate a trial move in the Monte Carlo procedure, a microstate 
$i$ is chosen at random, and a microstate $j$ that is a neighbor to microstate 
$i$ is also chosen at random.  If $i$ and $j$ do not belong to different 
macrostates, the trial move generation is repeated until they do, which is considered 
as one Monte Carlo step. 
The macrostate assignment of $j$ is then changed to match the macrostate 
of $i$, and the outside fraction $f^{out}_{trial}$ is calculated 
for this new definition of the states.  If $f^{out}_{trial} $$<$$ 
f^{out}$, we accept the new partitioning, and otherwise we accept 
it with Metropolis probability $min[1, exp^{(-\beta (f^{out}_{trial}-f^{out})}]$, 
where $\beta$ denotes inverse temperature.  
In this work, we use a very simple simulated annealing schedule in which 
$\beta$ is equal to $\frac{i}{10000}$, where $i$ is the Monte Carlo step number. 
We perform 5,000 Monte Carlo steps in each application of the method 
and check for the convergence of $f^{out}$. Since the method generally 
improves the state boundaries at each MC iteration, we determine the new dephasing times
for the states using the Anderson-Darling procedure after every 200 Monte 
Carlo steps. This procedure gives an ``on-the-fly" estimate of the 
dephasing times and also minimizes the correlated 
event time in a way that is more specific to the dephasing times. 
%{\bf (might be good to tell how many iterations were done, or how we decided it was converged)}
As a final step, after this Metropolis walk is converged, we restart 
the procedure using a very low temperature, so that essentially no 
uphill steps are accepted, and stop this procedure after 5000 Monte Carlo steps. 
%{\bf (perhaps also tell here how many iterations were done, or how we decided it was converged)}
Although the method should be applicable to an arbitrary partitioning 
of macrostates, it is advisable to initiate the process with a set 
of states that do not require extremely long dephasing times.  For 
every system, we initiated 50 Monte Carlo runs starting from 
the same state decomposition but with a different random number seed, 
and we selected the partitioning with the minimum value of $f^{out}$. 
We note that this framework is similar to the one developed by Chodera 
et.al.~\cite{choderapart}, where the goal was to discover the kinetic 
metastable states by maximizing the metastability index of a partitioning 
into $M$ macrostates. The metastability $Q$ was defined as the sum 
of self-transition probabilities at a given lag time, $Q=\sum_{i=1}^{L} 
T_{ii}(\tau)$, where $L$ is the number of macrostates and $T_{ii}$ 
denotes the self-transition probabilities.  

\section{Procedure for comparing MSM and QSD-KMC} \label{MSMmethod}

In the results below for the biochemical systems, we employ states 
defined by the Perron Cluster-Cluster Analysis (PCCA)~\cite{pccaweber}  
and compare the QSD-KMC results  to those from a Markov State Model 
(MSM)~\cite{msm1}-~\cite{msm4}.  Details of the MSM and PCCA 
methodologies are given in the supporting information.  
The output of PCCA is a fuzzy clustering of microstates to macrostates. 
However, in this work, we always consider the crisp assignment of the microstates 
to the macrostates and avoid the fuzzy assignment altogether, as 
QSD-KMC provides dynamical evolution of states with well-defined boundaries.  
In the MSM framework, this is achieved by generating a synthetic 
``microstate" trajectory at a given lag time and mapping it onto 
the crisp macrostates (from here on, we will refer to this model 
as {\it microstate MSM}). In addition, we show the results from a 
Markov model where the transition probability matrix is constructed 
directly over the set of macrostates (this model will be referred 
to as {\it macrostate MSM}).  This allows us to make a direct comparison 
of QSD-KMC macrostate evolution with the results obtained from different 
Markov models.  For evaluating our state optimization approach, we 
calculate the implied timescales by analyzing the transition matrix 
constructed on the final set of macrostates obtained and compare 
them with the timescales obtained via diagonalization of the transition 
matrix built using the crisp PCCA states. In this work, we use the PyEMMA 
software package~\cite{pyemma} for MSM construction and analysis. 

For validation of long time scale dynamics, we calculate the macrostate 
probability evolution, i.e., the probability to be in state $j$ at time 
$\tau$ given the system was in state $i$ at time $0$, for both the QSD-KMC trajectories
and the MD trajectories for all the states. The precise definition 
of this function is
\begin{equation}
\label{dckmcval}
P(i, j, \tau) = \frac{\sum_{k=1}^{ntraj} \sum_{t=1}^{T^{'}_{k}-\tau} 
\delta(S_{k}(t) - i) \delta(S_{k}(t+\tau) - j)}{\sum_{k=1}^{ntraj} \sum_{t=1}^{T^{'}_{k}-\tau} 
\delta(S_{k}(t) - i)}, 
\end{equation} 
where $ntraj$ is the number of trajectories, $T^{'}_{k}$ is the length 
of $k^{th}$ trajectory and $S_{k}(t)$ is the state of the system 
in $k^{th}$ trajectory at time $t$.  
In constructing the probability evolution plots for QSD-KMC using 
Eq.~\ref{dckmcval}, we make no distinction between residence in a 
state where the trajectory remains longer than a dephasing time and 
transient residence in that state during a correlated-event sequence.  
\redc{We believe this gives a meaningful definition of the state occupancy 
as a function of time for the purpose of calculating this function.
However, we note that because the properties of a system 
passing through a state quickly are not necessarily the same 
as the properties of a system that has settled into that state, using 
this state probability evolution function weighted by the equilibrium 
macroscopic property of each state is not sufficient to compute a fully 
accurate time evolution of the macrosopic properties of the system. On the other hand, 
as mentioned in Section~\ref{briefoverview}, greater detail about the instance trajectory
can be stored if desired; e.g., the microstate as a function of time can be recorded.}

\section{Results and Discussion} \label{discussion}

In this section, \redc{we first discuss the core concepts of QSD-KMC method 
on two easily visualized systems: a one-dimensional three-well potential
and a nonequilibrium system consisting of a one-dimensional sloped sine-wave potential.  
We show that the QSD-KMC trajectories accurately reproduce results from long benchmark
MD simulations on the same systems.}
We then compare the dynamics obtained from QSD-KMC with the underlying MD dynamics for 
two different biomolecular systems, dialanine and villin headpiece, 
to demonstrate the practical applicability 
of QSD-KMC to extract long time scale dynamics and the minimization 
of the outside time as a state optimization method. These systems 
enable us to show the potential of the QSD-KMC method to generate 
arbitrarily accurate statistics at any time resolution in scenarios 
where the dynamics consists of strongly metastable (long-lived) and 
weakly metastable (short-lived) states, thereby giving rise to multiple 
slow and fast processes. Such an analysis validates the robustness 
of the method to generate correct long-time statistics  when the 
dynamics does not exhibit ideal Markovian behavior.  The results 
for these biomolecular systems are structured as follows.  First we 
define states based on MSM and PCCA. We calculate the probability 
evolutions using QSD-KMC,  microstate MSM (at two lag times) and 
macrostate MSM, and compare them to the results from the MD trajectories.  
Then we use simple rectangular state boundaries for the same number 
of states and evaluate the resulting probability evolutions using 
these three approaches. Finally, we apply the outside-time based 
state optimization method, starting from the simple rectangular state 
decomposition and show how well they reproduce the state boundaries 
obtained using PCCA.  

\begin{figure}
\centering     %%% not \center
\subfigure[A three-well potential function]{\label{three-well:a}\includegraphics[width=65mm]{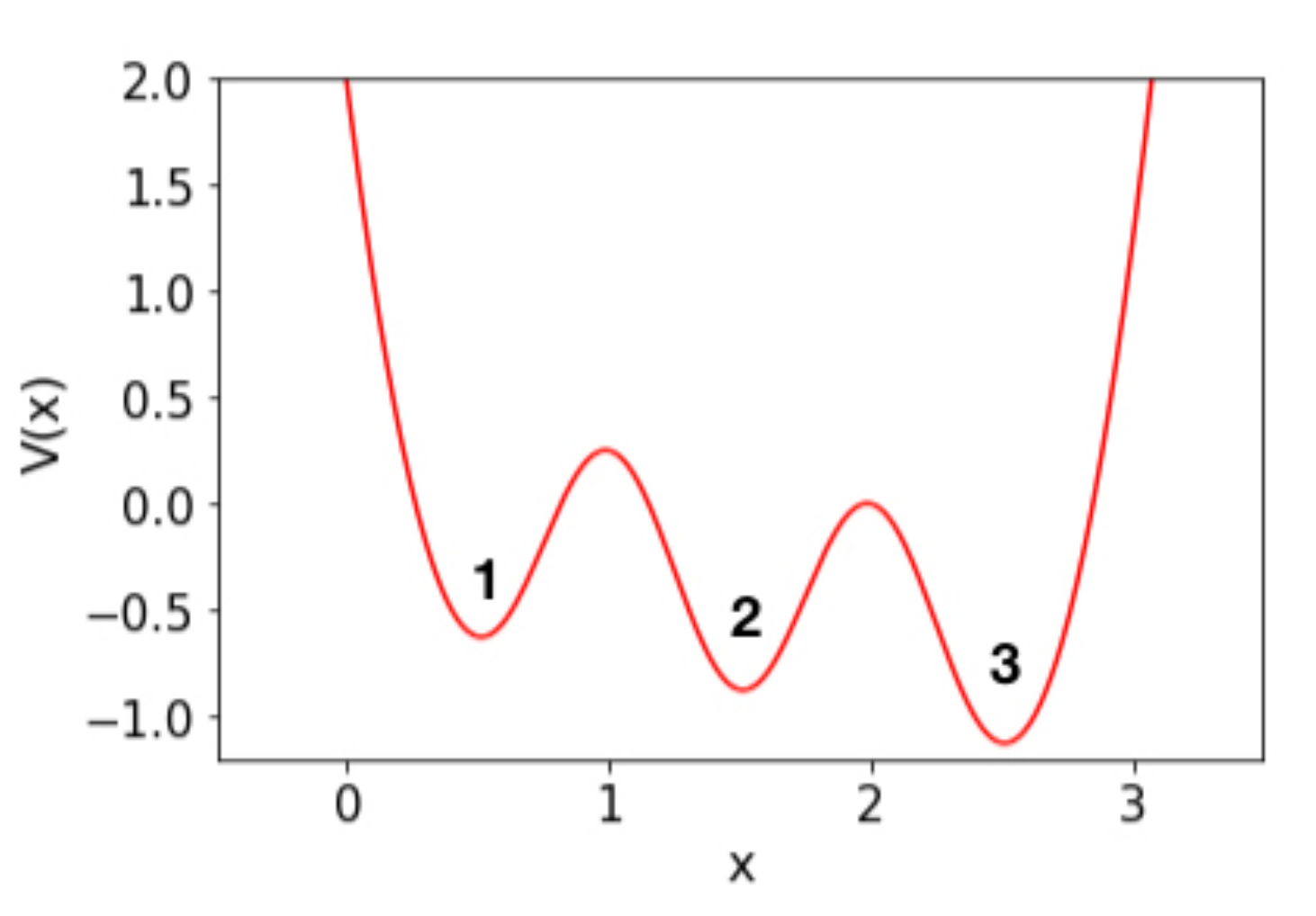}}
\subfigure[Survival probability function for state 2]{\label{three-well:b}\includegraphics[width=65mm]{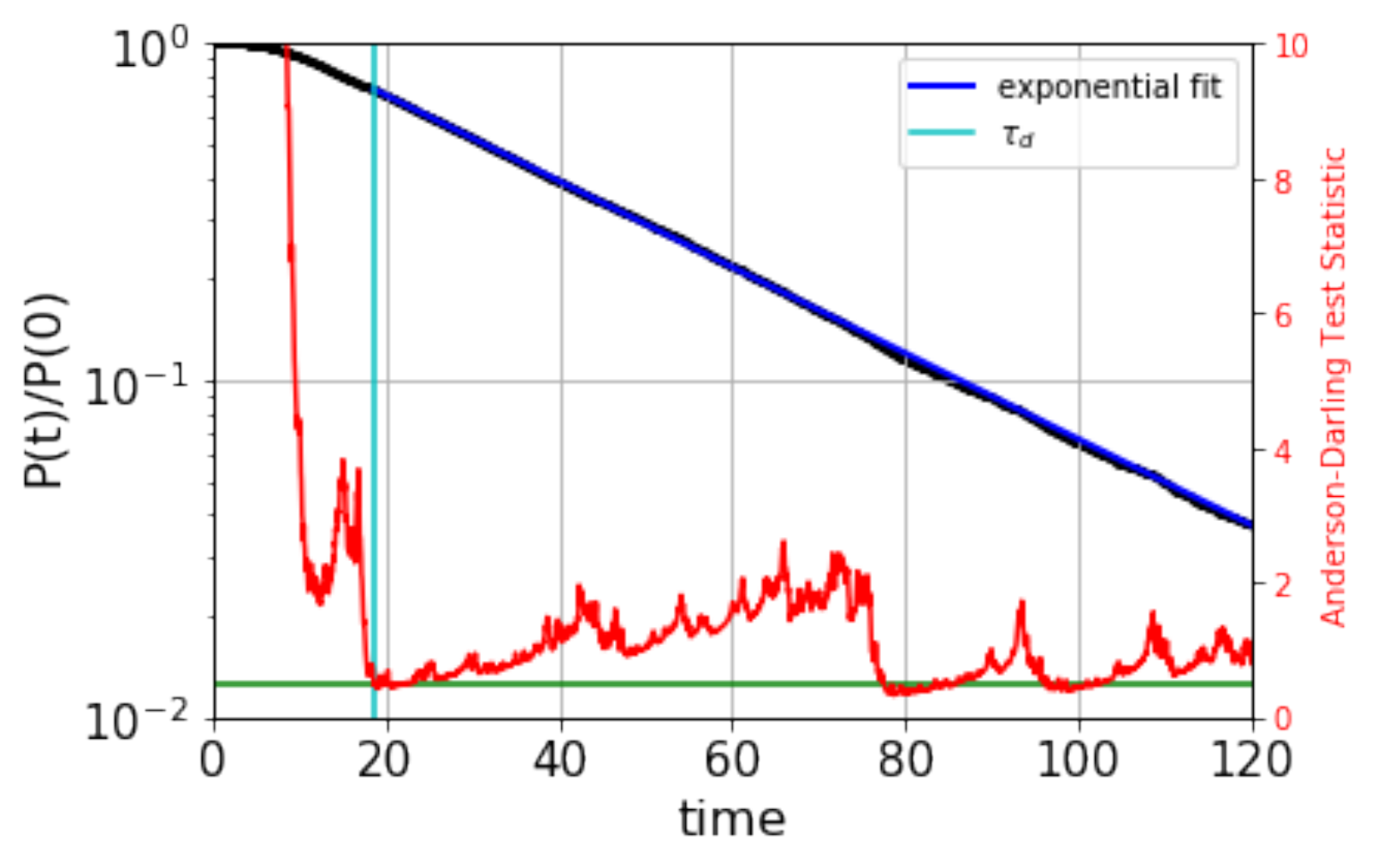}}
\subfigure[Different densities computed for state 2]{\label{three-well:c}\includegraphics[width=65mm]{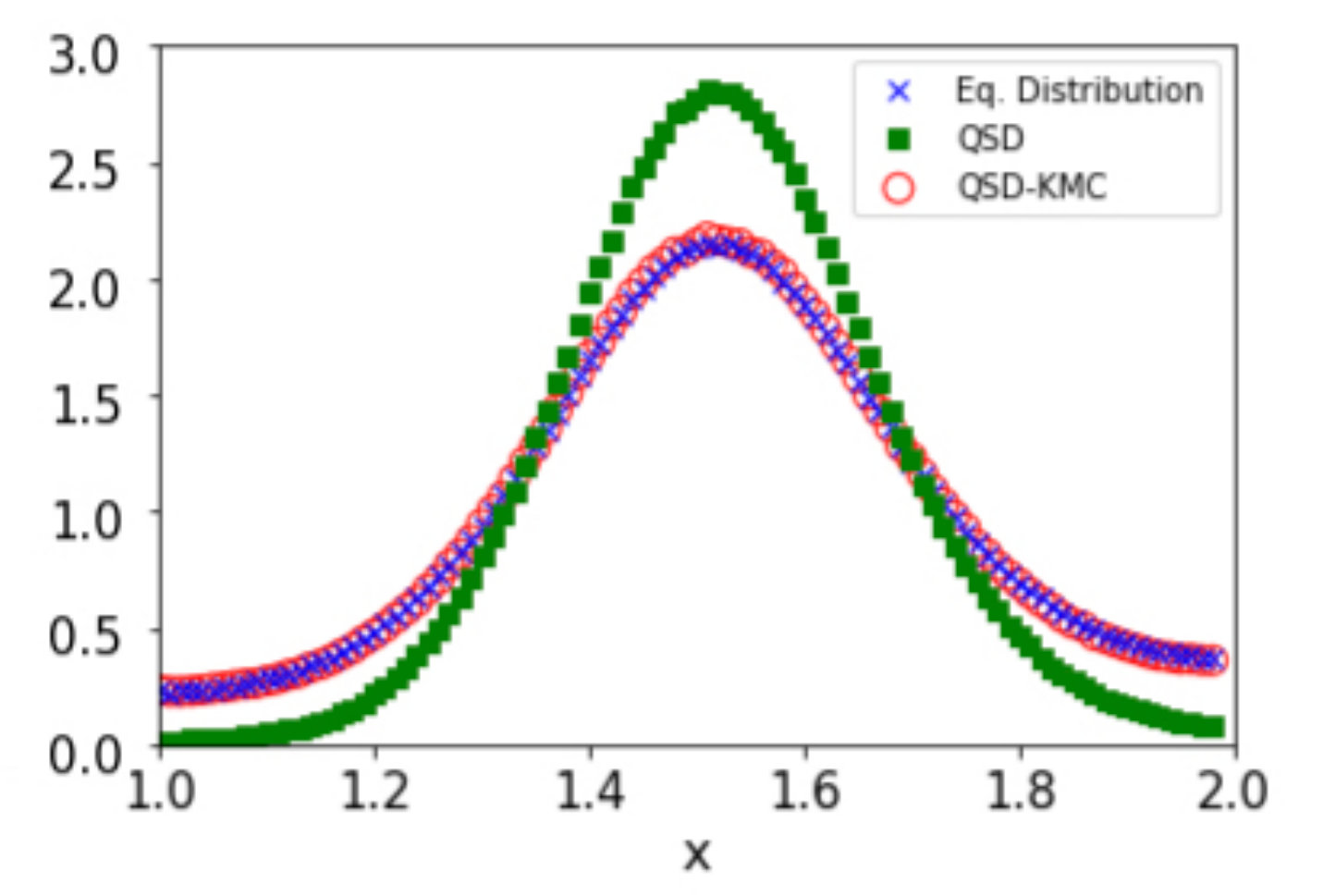}}
%\subfigure[]{\label{three-well:c}\includegraphics[width=65mm]{Figures/state_2_dephasing_time_distributions.png}}
\caption{\redc{(a) A three-well potential function. (b) Survival probability function and Anderson-Darling dephasing time for state 2.
(c) Different densities computed for state 2 : the Boltzmann (equilibrium) density, the QSD density and the density accumulated along the path of a QSD-KMC trajectory.}}
%{\bf Animesh, can you fix the c description, and also maybe plot using dotted lines instead of symbols for rho eq and qsd-kmc?}
\label{three-well}
\end{figure}

\bigskip
\noindent
\redc{
\textbf{One-dimensional potential -- equilibrium system}

We first consider dynamics in a one-dimensional, three-well potential.
%%%This allows us to demonstrate and discuss some core concepts on an easily 
%%%visualized system. 
%%%{\bf we can take this out again if you want, but for now I am voting to leave it in}
This allows us obtain results to very high 
precision, making the accuracy of the method very clear.
The potential, shown in Figure~\ref{three-well:a}, is a sloped cosine curve with 
a curvature-matched harmonic wall on the left side of the lefthand 
state (state 1) and on the right side of the righthand state (state 
3),  
%
%\begin{equation}
%\label{one-dim-3state}
%V(x) =  {\bf need \ equation \ here}
%\end{equation} 
%
\begin{equation}
\text{V(x)} =
\begin{cases}
 \frac{1}{2} \cos(2\pi x) + sx , \ \ \ \ \frac{1}{2} \leq x\leq \frac{5}{2}
\\
 \frac{1}{2} \cos(2\pi \left(\frac{1}{2}\right)) + sx , \ \ \ \ x < \frac{1}{2}
 \\  
 \frac{1}{2} \cos(2\pi \left(\frac{5}{2}\right)) + sx , \ \ \ \ x > \frac{5}{2}, 
\end{cases}
\end{equation}
where $s=-1/4$ is the slope. 
In this system, the three states are very similar other than being 
at different energies.  We use a fairly low friction so that double-jump 
correlated events are more likely to occur.  
We integrate the Langevin equations of motion using the BAOAB method~\cite{boab},
using $k_{b} T$=0.5, friction $\gamma$  = 0.1 inverse time units, time step $\delta t$ = 0.05 and mass $m$ = 1.

The benchmark simulation providing the correct answer is a single, 
long MD simulation, of length 2,000,000 time units.
To generate the QSD-KMC model, 10,000 short trajectories were initiated 
in each of the three states; each trajectory was continued until it 
escaped from its state. This corresponds to the first stage of the 
two-stage procedure described in Section~\ref{corr-instance}.  We determined the 
dephasing time for each state using the Anderson-Darling procedure 
described in Section~\ref{choice}, as shown in Figure~\ref{three-well:b} for state 2.  
There is clearly a significant deviation from exponentiality at times 
shorter than 20 time units, after which the Anderson-Darling statistic 
rapidly drops down to the 0.5 threshold.  Applying this dephasing 
time, roughly 25\% of trajectories were discarded from each 
state, and the QSD escape times were determined from the average
escape times, offset by $\tau_d$, of the remaining trajectories.
Then, following the second stage of the procedure, these remaining 
trajectories were continued forward until they dephased again in some state,
each of these representing an instance trajectory. 
\begin{figure}
\centering     %%% not \center
\includegraphics[width=1.1\columnwidth]{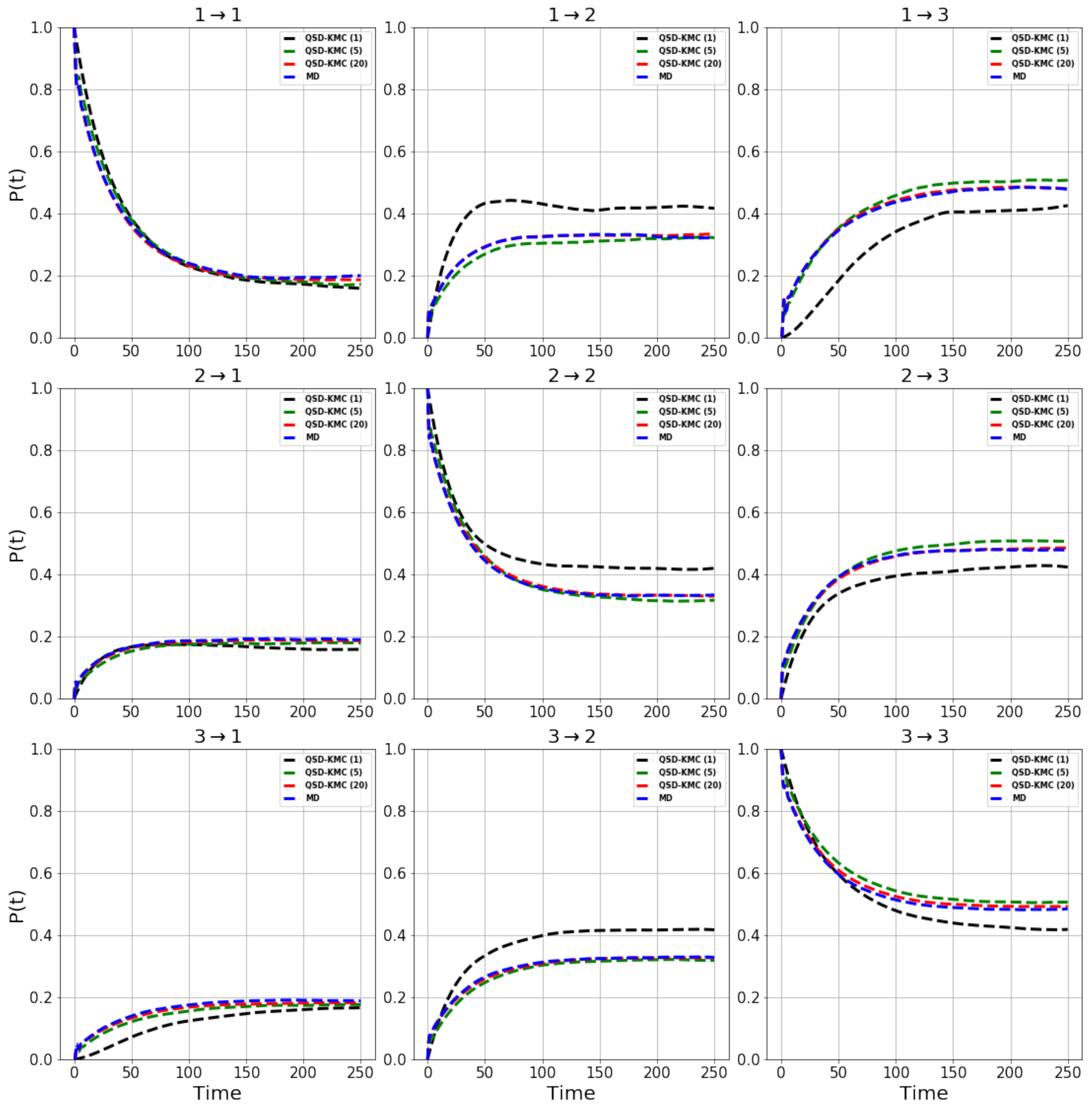}
\caption{\redc{State-to-state probability evolutions calculated from QSD-KMC 
and from the benchmark MD trajectory
for the three-state system. 
The dephasing times of 20 time units for all the states are estimated 
from the Anderson-Darling procedure described in Section~\ref{choice}. 
For comparison, we also show the QSD-KMC results using $\tau_{d}$ = 1 and 
5 time units.  
}}
\label{ck-three-well}
\end{figure}
\begin{figure}
\centering     %%% not \center
\includegraphics[width=0.5\columnwidth]{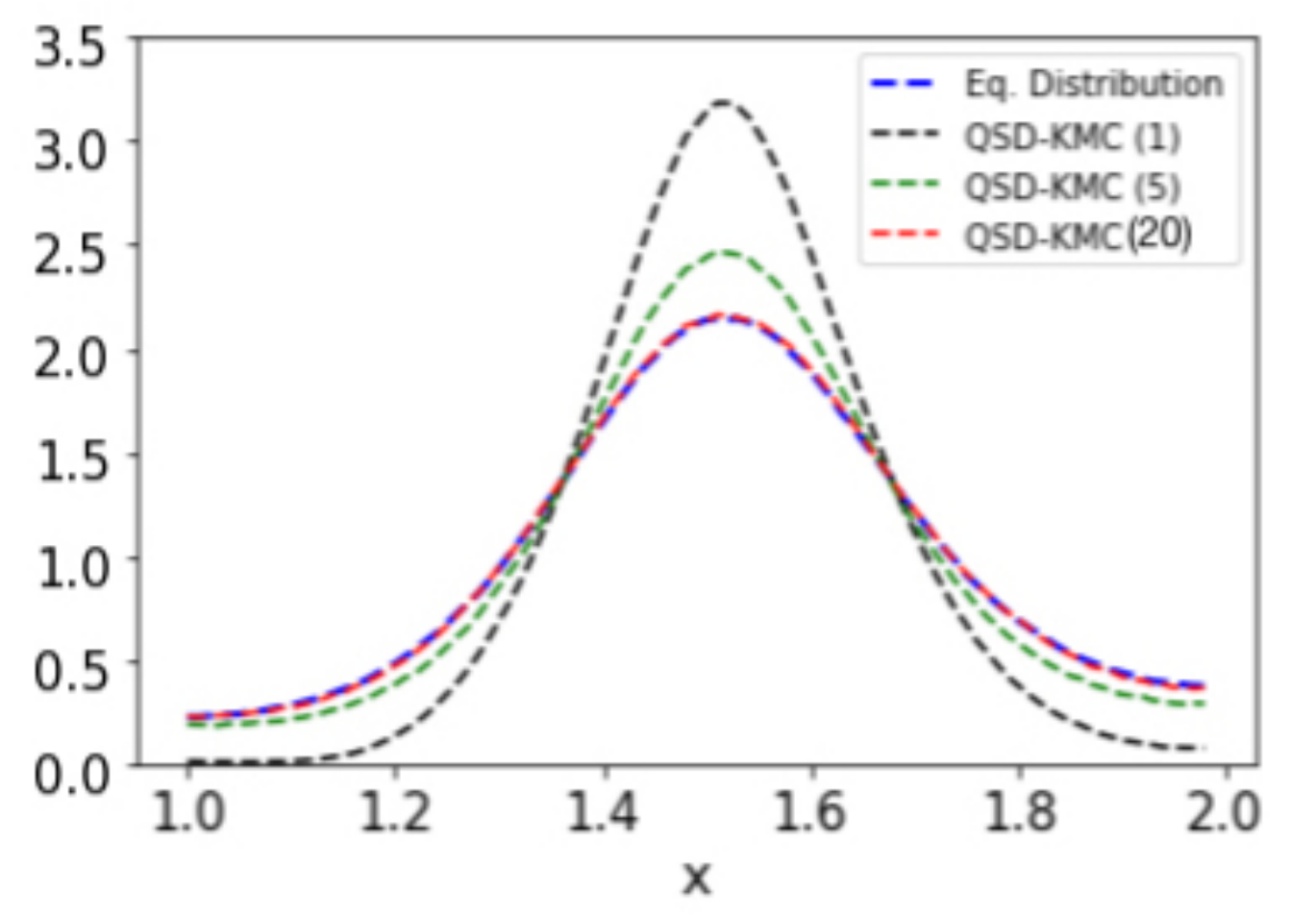}
\caption{\redc{QSD-KMC density as a function of the dephasing time (number of time 
units shown in parenthesis) for the three-state system.}} \label{three-well-dephasing}
\end{figure}
Figure~\ref{three-well:c} shows the QSD density for state 2, generated 
by sampling points from trajectories that have remained 
at least a time $\tau_{d}^{(2)}$ in the state.  (The QSD densities for states 
1 and 3 are very similar to this.) The QSD density, $\rho_{QSD}$, is clearly different than 
the equilibrium (Boltzmann) distribution, $\rho_{eq}$, which was 
generated using a Metropolis Monte Carlo procedure, rejecting steps 
that attempted to leave the state (i.e., corresponding to reflecting 
boundary conditions).  The QSD is more peaked in the center of the 
state, and has less density than $\rho_{eq}$ near the boundaries. 
This is a general characteristic of QSDs, caused by the absorbing 
boundary condition at the edge of the state.  Some trajectory points 
near the boundary that would contribute to the Boltzmann density 
are missing from the QSD density because near the boundary there 
is a higher probability that the trajectory entered the state recently, 
more recently than a dephasing time.  For very deep 
states, $\rho_{QSD}$ approaches the shape of $\rho_{eq}$, because 
the potential energy is so high near the edge of the state that further 
depletion by the absorbing boundary is hardly noticeable.  

Figure~\ref{ck-three-well} shows the state-to-state probability evolution 
calculated from the benchmark MD trajectory and from the QSD-KMC 
trajectories at different dephasing times.  The QSD-KMC curves (at 
$\tau_{d}$ = 20 time units) are seen to agree extremely well with 
the direct MD result; however for a short dephasing time, the results from 
QSD-KMC disagree with the MD results, signifying the importance of 
the dephasing time.  

Because of the simplicity of this system, it was easy to store the 
trajectory position at every MD step along each instance path (i.e., 
from the instant the trajectory left its initial state until the 
time it finished dephasing in the final state).  This is in contrast 
to the biochemical systems we present below, for which only the macrostate 
information was stored (although it would be easy to store the
microstate information along the instance path as well).  Using this 
detailed information, we have a model that can predict 
the evolution essentially continuously in time.  
For example, in Figure~\ref{three-well:c}, 
we can see that the density accumulated along the path of a long QSD-KMC 
trajectory gives a precise match to the Boltzmann density, as it should
for a system in equilibrium. 
% xxx - changed it a bit more -- see what you think. -Art
To compute this QSD-KMC density, the trajectory positions along the 
QSD-KMC path are stored and binned into a histogram, and the histogram 
is normalized for each state.  This QSD-KMC density is comprised 
of contributions from trajectories that pass through the state without 
settling (e.g., as part of a double jump or longer jump), as well 
as contributions from the QSD density for time durations when the 
system is in the QSD and has not yet reached the next escape time.  
In Figure~\ref{three-well-dephasing}
we show how the density obtained from a QSD-KMC
trajectory varies with the dephasing time. It can be clearly seen 
that the distributions 
constructed using a very short dephasing time disagree with the MD results, showing 
that the information from the MD data is not simply reused; establishment 
of the QSD within the states plays a significant role in generating 
accurate statistics.  
%{\bf OPTIONAL - RIGHT HERE, it would be nice to be able to show some sort of time correlation
%function like we have talked about.  But if it is a big pain to generate it,
%don't worry about it.}

%{\bf not sure if we will mention the following, but we could (we will 
%definitely show it for the nonequilib case):}
\begin{figure}
\centering     %%% not \center
\subfigure[A sloped sine wave potential function]{\label{stair-well:a}\includegraphics[width=65mm]{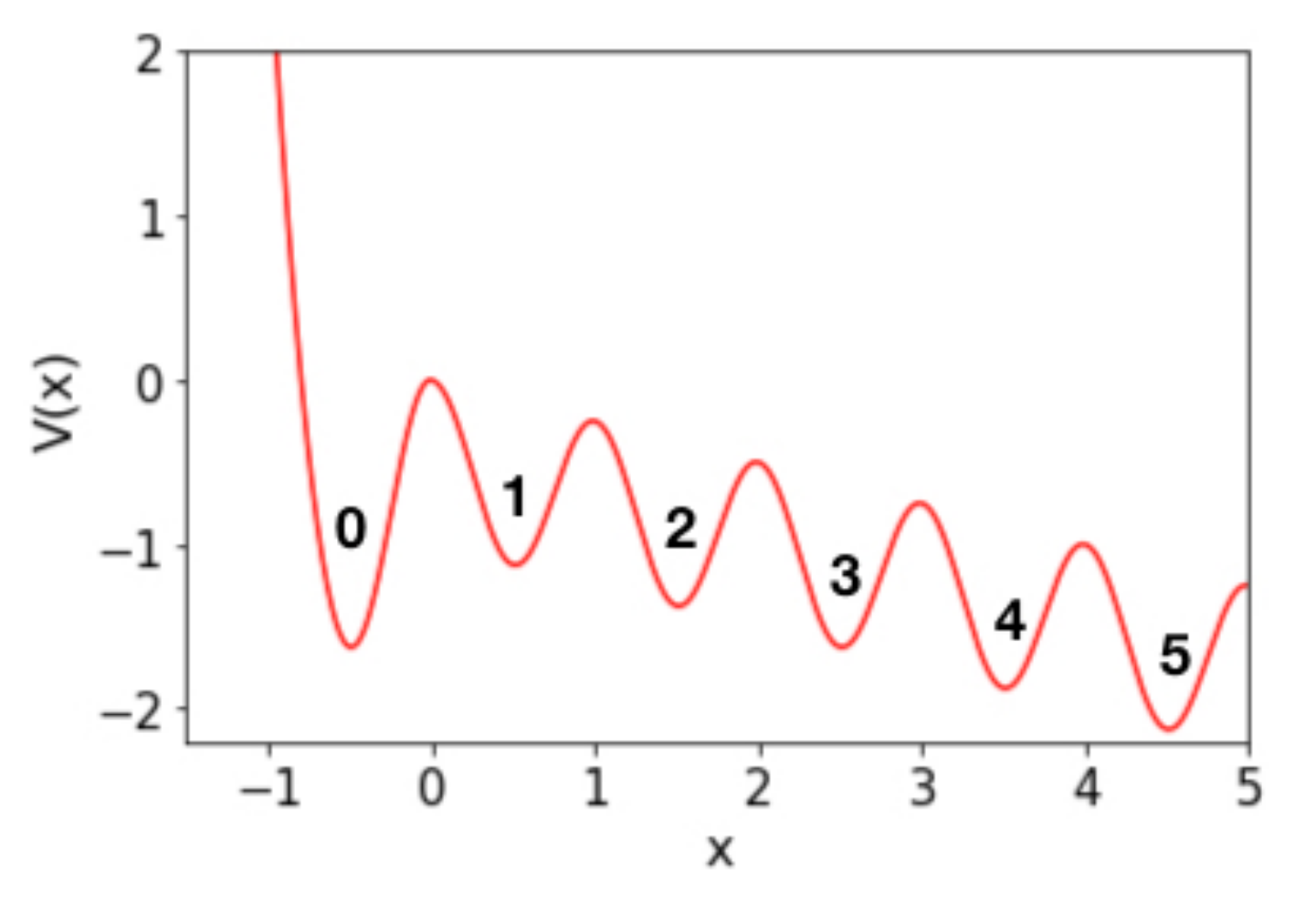}}
\subfigure[Survival probability function for state 1]{\label{stair-well:b}\includegraphics[width=65mm]{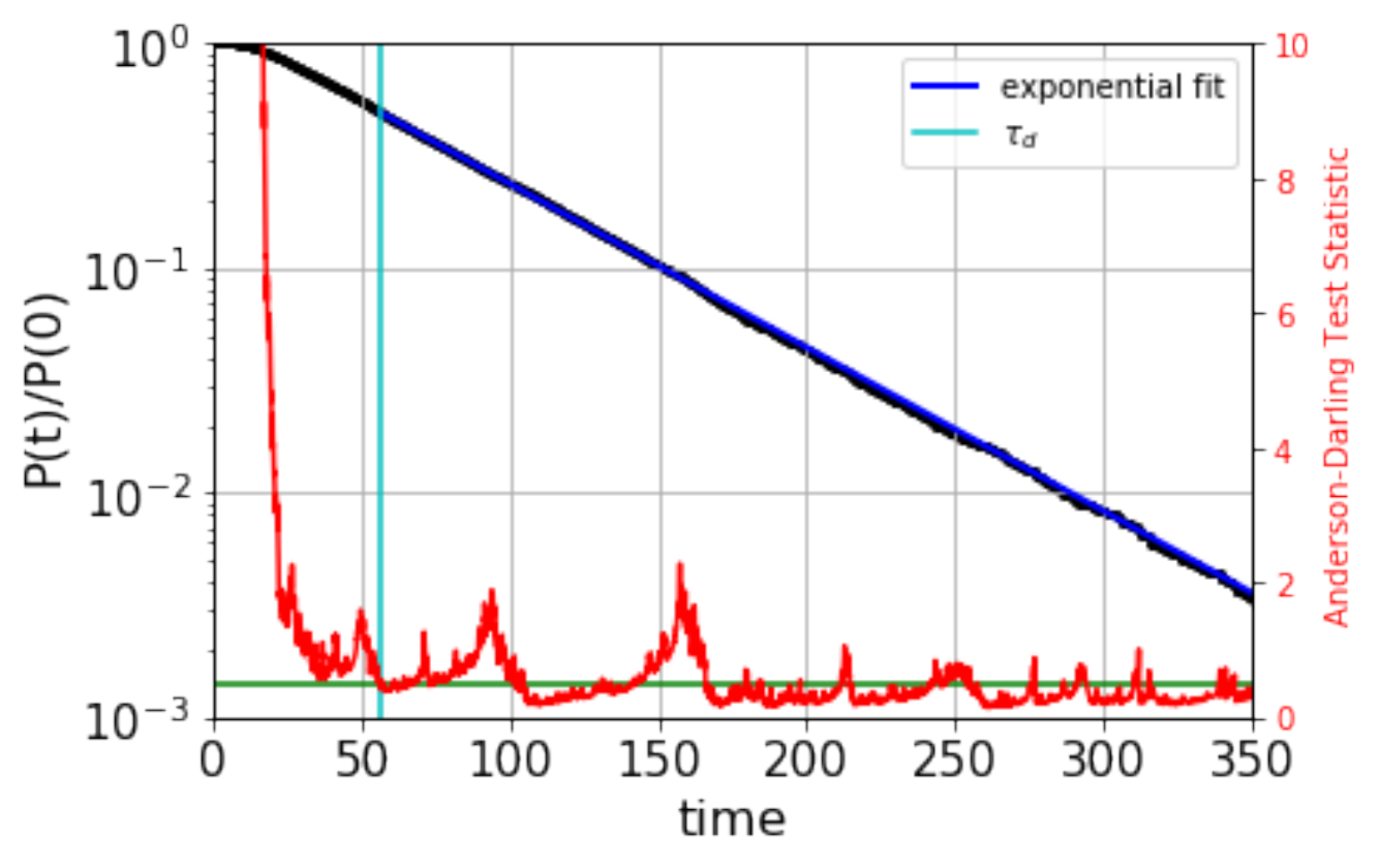}}
\subfigure[Different densities computed for state 1]{\label{stair-well:c}\includegraphics[width=65mm]{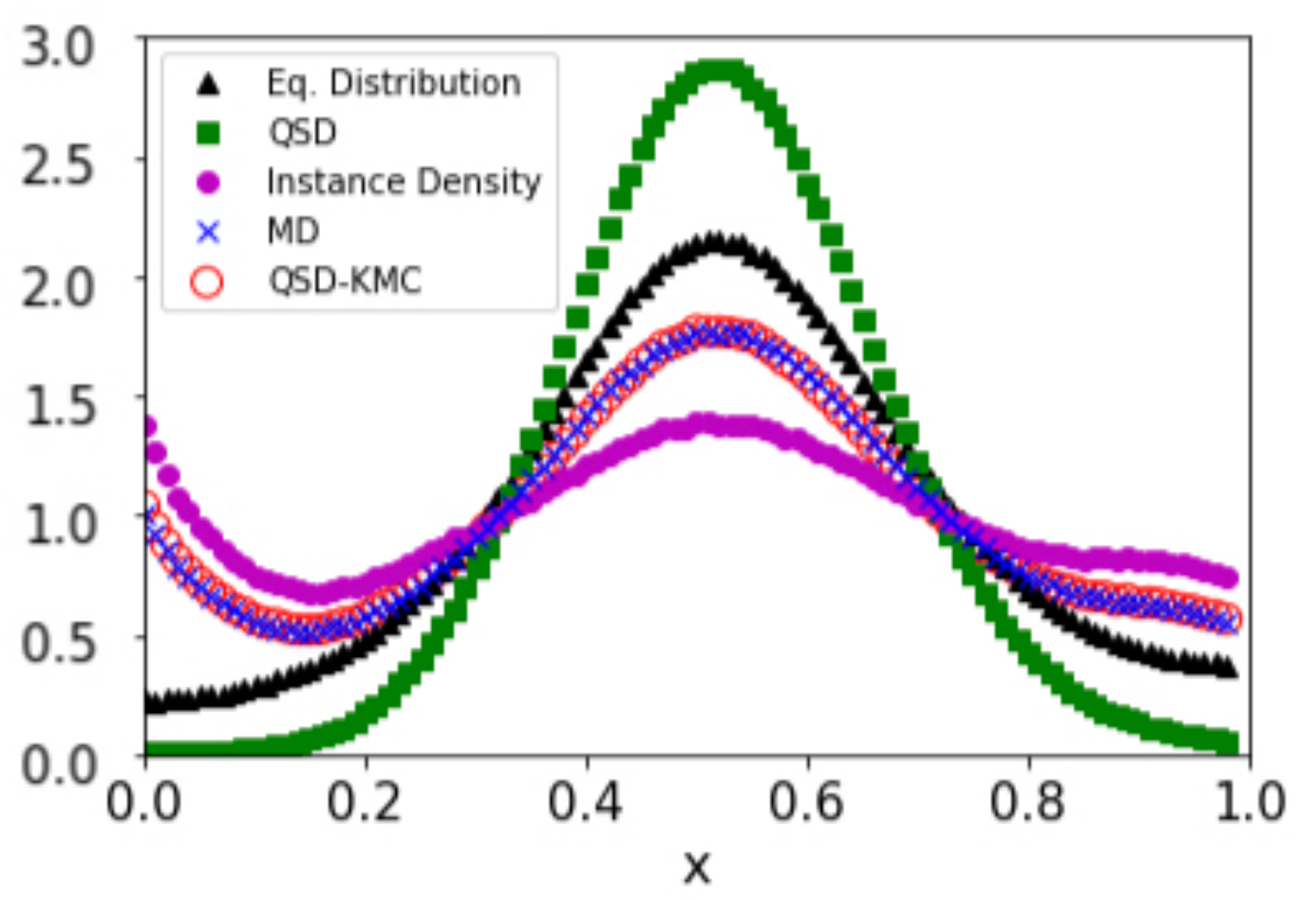}}
\subfigure[QSD-KMC density as a function of the dephasing time]{\label{stair-well:d}\includegraphics[width=65mm]{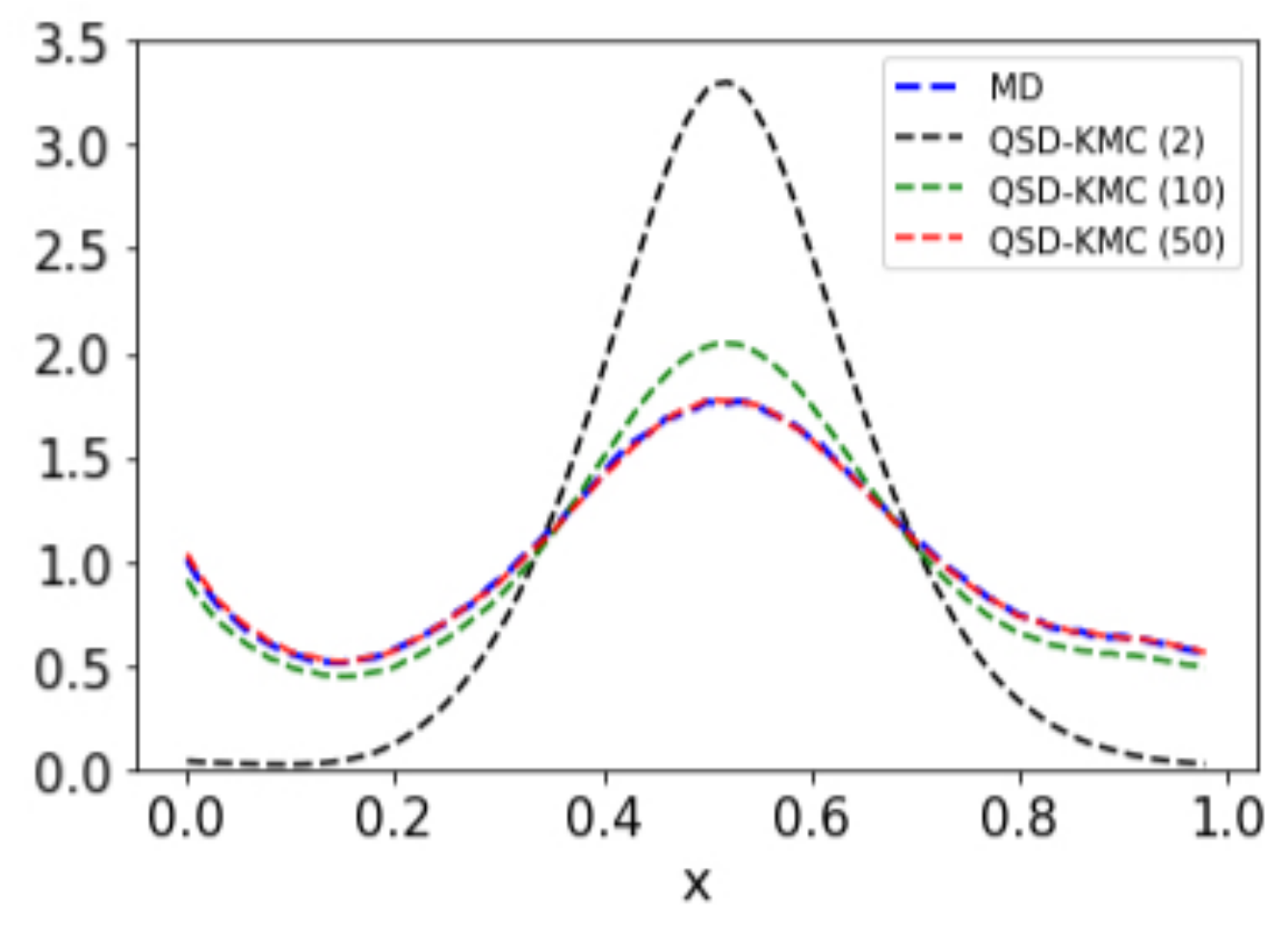}}
\subfigure[Different densities computed for state 2]{\label{stair-well:e}\includegraphics[width=65mm]{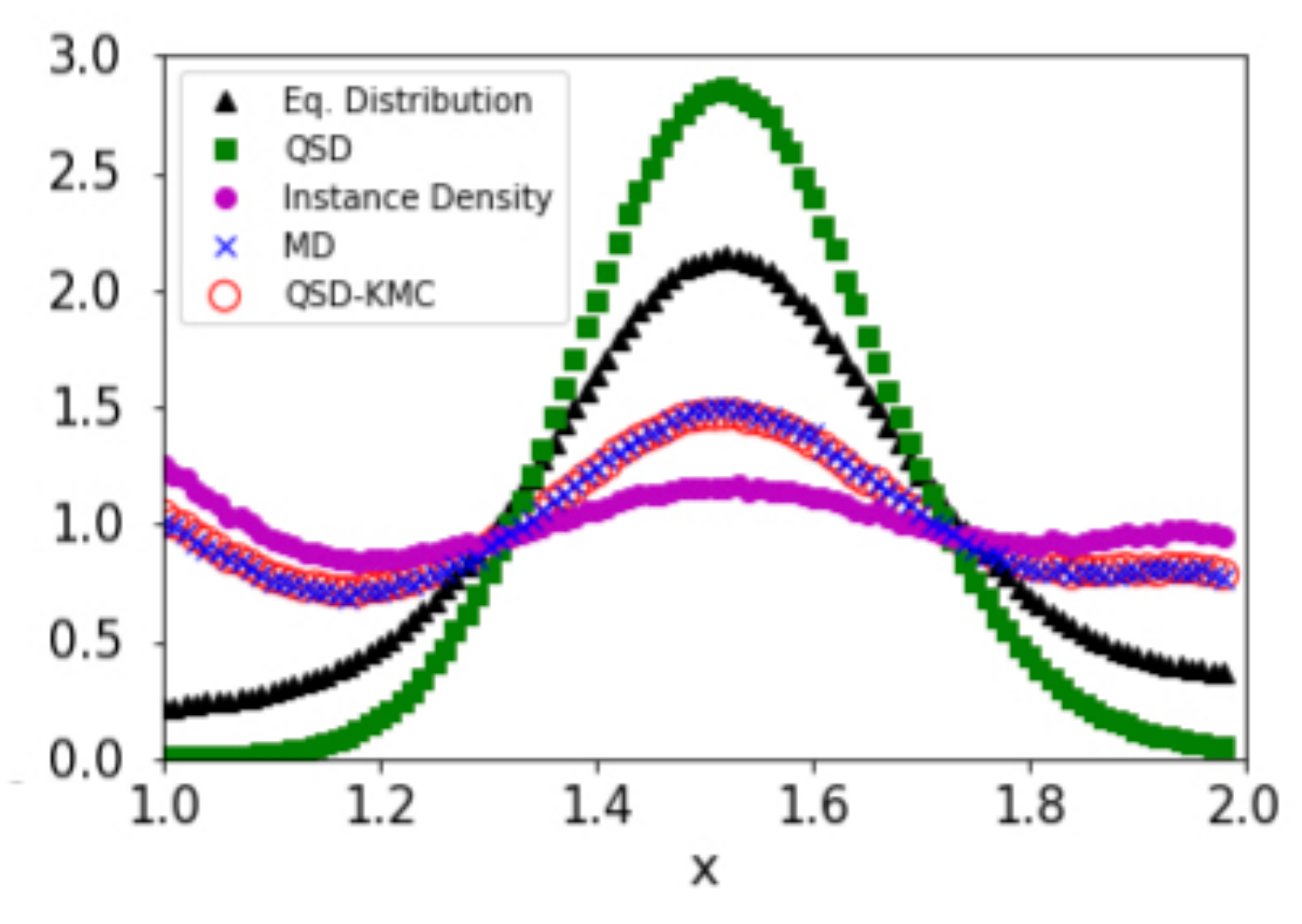}}
%\subfigure[]{\label{three-well:c}\includegraphics[width=65mm]{Figures/state_2_dephasing_time_distributions.png}}
\caption{\redc{(a) A sloped sine wave potential function. (b) Survival probability function and Anderson-Darling dephasing time for state 1.
(c) Different densities computed for state 1: the Boltzmann (equilibrium) density, the QSD density, the density contribution 
from the pass-through trajectories, the (non-equilibrium) density obtained from the benchmark MD
trajectories and the density accumulated along the path of a QSD-KMC trajectory.  (d) QSD-KMC density as a function of the dephasing time (number of time 
units shown in parenthesis) for state 1. (e) Different densities computed for state 2 similar to the state 1 case.}}
\label{stair-well}
\end{figure}

\bigskip
\noindent
\textbf{One-dimensional potential - nonequilibrium driven system}

Next we consider driven, nonequilibrium dynamics in the one-dimensional 
sloped cosine-wave potential given by
\begin{equation}
\text{V(x)} =
\begin{cases}
 \frac{1}{2} (\cos(2\pi x) - 1) + sx , \ \ \ \ x > 0
\\
 \frac{7}{8} (\cos(2\pi x) - 1) + sx , \ \ \ \ -1/2 \leq x \leq 0
 \\  
\frac{7}{8} (\cos(2\pi \left(-\frac{1}{2}\right))- 1) + sx , \ \ \ \ x < -1/2, 
\end{cases}
\end{equation}
where $s=-1/4$ is the slope.
This potential is shown in Figure~\ref{stair-well:a}.
The exact MD benchmark for this system is computed by initiating 
trajectories in state 0, the highest-energy (and a relatively deep) 
state.  After equilibration in state 0, each trajectory escapes from 
this state and falls down through the staircase of wells;  it loses 
its memory and equilibrates in some wells, but passes quickly through 
others.  There is no end to the staircase, so equilibrium is never 
achieved.  A trajectory may hop back up the staircase in some cases, 
but it is more likely to escape in the forward $x$ direction.  Because 
we employ a very low Langevin friction, there is a high probability 
that it will simply pass through a state as part of a correlated 
multiple-jump event, and for a trajectory moving in the $+x$ direction, 
this probability increases with the length of the jump, as it tends
to pick up speed at this low friction.  
(Ferrando et al~\cite{fer3} have directly studied the friction 
dependence of the jumping behavior in a
tilted cosine system).
This is clearly non-Markovian dynamics, and we demonstrate 
the accuracy of the QSD-KMC method, again implemented using many 
short trajectories, by comparing the predicted steady-state density 
in $x$ for this nonequilibrium system with the exact MD benchmark.  

The integration method and parameters are the same as in the equilibrium 
three-state system, except that the friction is lower ($\gamma$ = 0.05 
inverse time units).  To generate the QSD-KMC model, 100,000 short 
trajectories were initiated in each of states 0 through 4, and 10,000 
trajectories were initiated in each of states 5 through 10.  The 
two-stage procedure to determine the dephasing times, the QSD escape 
times, and the instance trajectories was then carried out just as 
in the three-state model above, except that any trajectory that reached 
state 11 was terminated, on the assumption that it had a negligible 
chance of returning to state 2 or 1, the states we focus on in our 
analysis.  For the long-MD benchmark results, we averaged over 100,000 
trajectories that were initially equilibrated for 50 time units 
in state 0.  The termination at state 11 was again imposed.  

The Boltzmann probability density for state 1 is shown in Figure~\ref{stair-well:c}.
Even though this is a nonequilibrium system, this Boltzmann 
density is exactly the same as for the equivalently shaped middle 
state (state 2) of the three-state equilibrium system.  If the friction 
setting were the same, this would also be true of the QSD density, 
the Anderson-Darling behavior, and the dephasing time.  However, 
because we are using a lower friction, the QSD density changes and the
dephasing time increases to 55 time units, compared to 20 time units 
for the three-state system, as can be seen in Figure~\ref{stair-well:b}.  
As in the three-state case, we store the trajectory position at every 
time step along each instance trajectory segment, making it possible 
to generate the density accumulated along the QSD-KMC trajectory 
path to high precision in $x$.

Focusing on state 1, Figure~\ref{stair-well:c} shows that the QSD-KMC 
density, agrees essentially perfectly with the exact 
MD results.  QSD-KMC is giving the correct nonequilibrium 
density for this system at every point in space.  Because this is 
a nonequilibrium system, the density in state $i$ is a function 
of both the shape of the potential within state $i$, as well as shapes 
and connectivities of the states outside state $i$.  The exact density 
%%%($\rho_{exact}$) % we never use this, so deleting...
turns up towards the left edge of the state, differing 
significantly from both the equilibrium density and the QSD density.  
This higher density near the left edge results from trajectories 
that pass through state 1 without settling (98\% of trajectories 
do this).  At this very low friction, such trajectories pass through 
in a nearly energy conserving way, moving more slowly where the potential 
V is high, and faster where V is low.  They hence spend more time 
(contributing more to the density) on the left edge of the state.  
The normalized density contribution from just these pass-through 
trajectories, (labeled ``instance density") is shown in 
Figure~\ref{stair-well:c}, and is seen to dominate the shape of the 
QSD-KMC density on the lefthand side.  

We see again that using a proper dephasing time is important, as using 
a smaller value gives a very different result for the QSD-KMC density, 
as can be seen in Figure~\ref{stair-well:d}, and this is even more 
important at this lower friction setting.  The QSD-KMC density is 
well converged by $\tau_d$=50, in basically perfect agreement with 
the density from the long-MD benchmark (as was also shown in Figure 
\ref{three-well-dephasing}). The direct MD benchmark run on the nonequilibrium 
system is, by definition, the same as using a dephasing time of infinity, 
because with $\tau_{d}$=$\infty$, no trajectory would ever settle, 
and everything would be made up from the instances that start in 
state 0. In contrast, using a very short dephasing time causes the density 
to look very much like $\rho_{QSD}$, because almost every trajectory 
entering the state ``dephases'', so the QSD-KMC trajectory spends most of its time
in the QSD for this state.  
\begin{figure}
\centering     %%% not \center
\includegraphics[width=1.1\columnwidth]{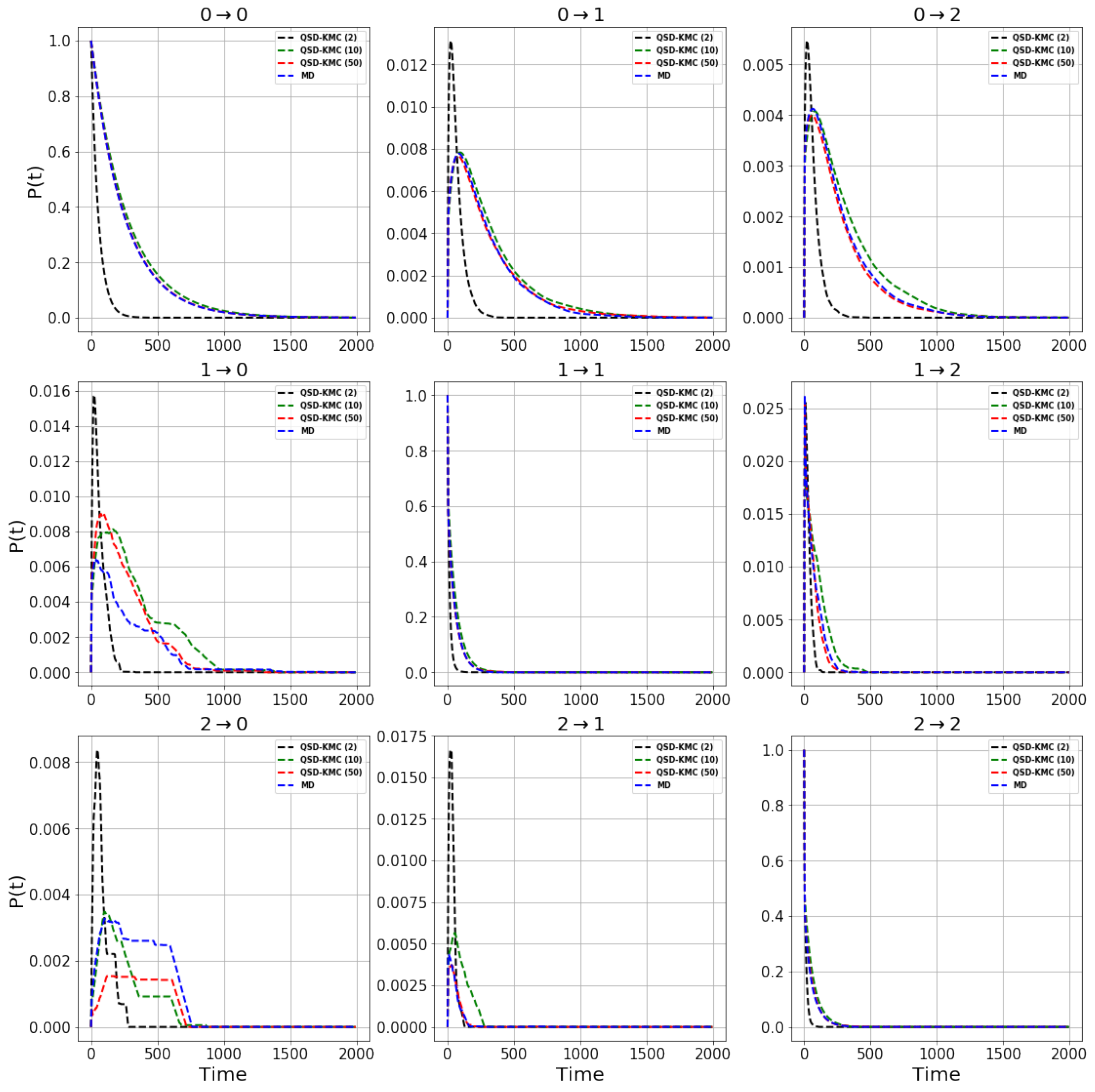}
\caption{\redc{State-to-state probability evolutions calculated from QSD-KMC 
and the benchmark MD trajectories for the non equilibrium one-dimensional system. 
The dephasing time of 50 time units for all the states are estimated 
from the Anderson-Darling procedure described in Section~\ref{choice}. 
For comparison, we also show the QSD-KMC results at $\tau_{d}$ = 1 and 
5 time units.}  
}
\label{stair-well-ck}
\end{figure}

Figure~\ref{stair-well:e} shows these same quantities for state 2.
The peak height of the steady state distribution is seen to be 
lower in state 2 than it is in state 1, which we can understand 
as follows.  As just discussed, a significant fraction of the trajectories 
entering state 2 from state 1 passed quickly through state 1, and picked
up some speed.  For this type of trajectory, the probability of settling in state 2
is smaller than it is for a trajectory that was settled in state 1 before entering state 2.
The effect of this, which is to lower the center peak in the QSD-KMC density in state
2 relative to state 1, can be seen by comparing Figure~\ref{stair-well:e} 
and Figure~\ref{stair-well:c}.  Again, the density predicted by the QSD-KMC
model is in extremely good agreement with the density from the exact-MD benchmark.

Figure~\ref{stair-well-ck} shows the state-to-state probability evolutions 
computed from the QSD-KMC trajectory using different dephasing times.  We 
note that because this is a nonequilibrium system with an infinite 
number of states, the probability values at a given time across a 
row of plots (e.g., 0$\rightarrow$0,  0$\rightarrow$1, and 0$\rightarrow$2) do not sum to unity, as 
they do for the equilibrium case (e.g., Figure~\ref{ck-three-well}).  Rather, at 
times greater than $t$=1500, the probabilities have all decayed to 
essentially zero, because the trajectories have passed through this 
region, never to return.  Again, the agreement with the MD benchmark 
results is essentially perfect.  Here, the scales for the different 
plots are chosen to allow the curves to be seen even when the values 
are very small, so the apparent disagreements between the QSD-KMC 
curve with $\tau_d$=50 and the MD curve for cases 1$\rightarrow$0 and 2$\rightarrow$0 are 
actually extremely small. }

%%********** end of the nonequilibrium staircase system section ************

%%*********************************
%% Dialanine
%%*********************************

\bigskip
\noindent
\textbf{Dialanine}: 

We study the long timescale dynamics of 
dialanine in explicit water. Dialanine is extensively used as a simple 
system to demonstrate important concepts in biologically relevant 
methodologies since it is a well-studied system~\cite{animesh} and 
the conformational changes can be characterized in terms of its backbone 
dihedral angles (Ramachandran angles, $\phi$ and $\psi$).  We generate 
fourteen 1-$\mu$$s$ long MD trajectories, saving snapshots at 2-$ps$ 
intervals. The technical details of the simulations are reported 
in the supporting information.  Figure~\ref{ala-free:a} 
shows the free energy landscape of dialanine in the $\phi-\psi$ dihedral 
angles.  
\begin{figure}
\centering     %%% not \center
\subfigure[Free energy map of dialanine in $\phi-\psi$ space]{\label{ala-free:a}\includegraphics[width=65mm]{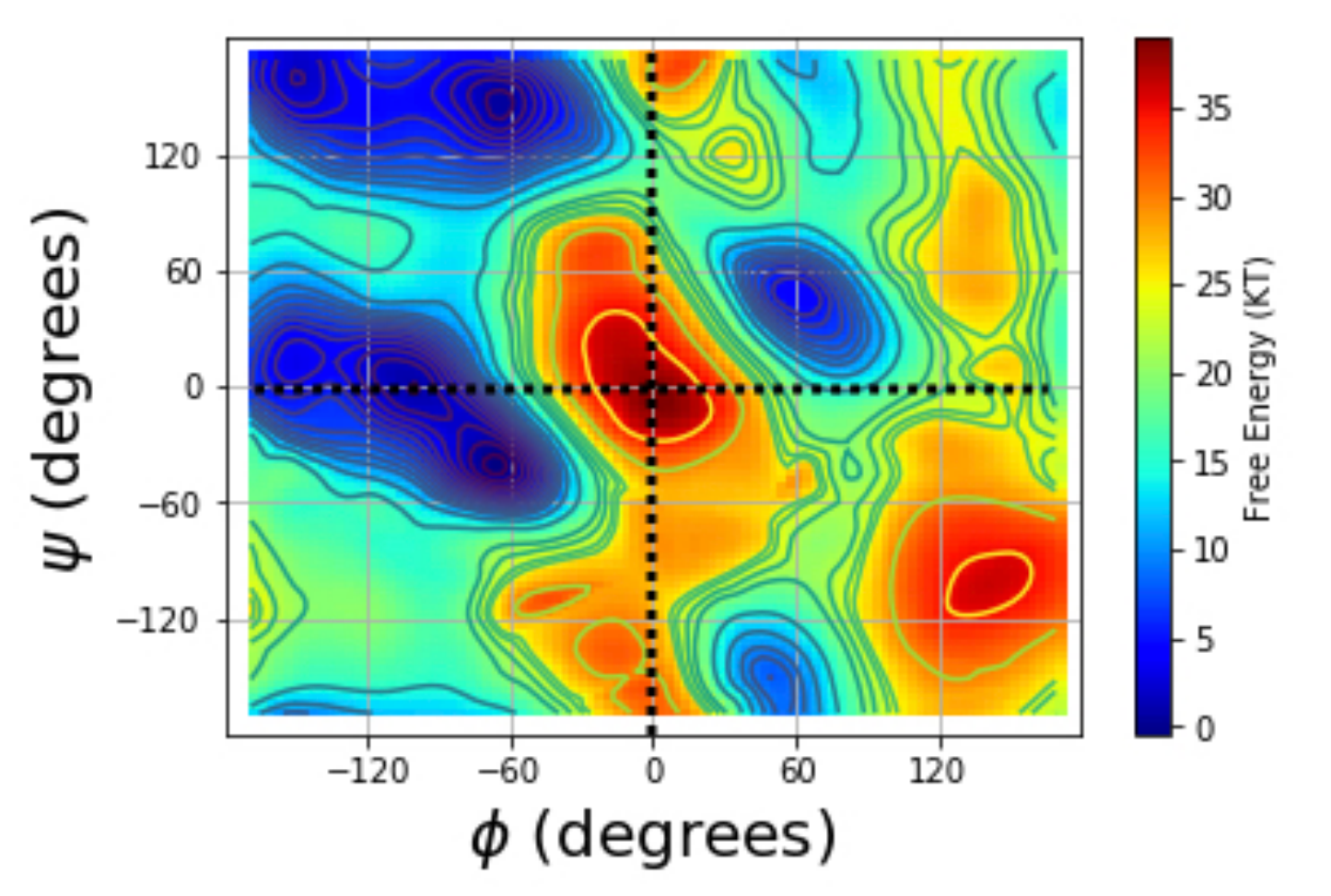}}
\subfigure[Relaxation timescales calculated as a function of lag time for 
dialanine in explicit water]{\label{ala-free:b}\includegraphics[width=65mm]{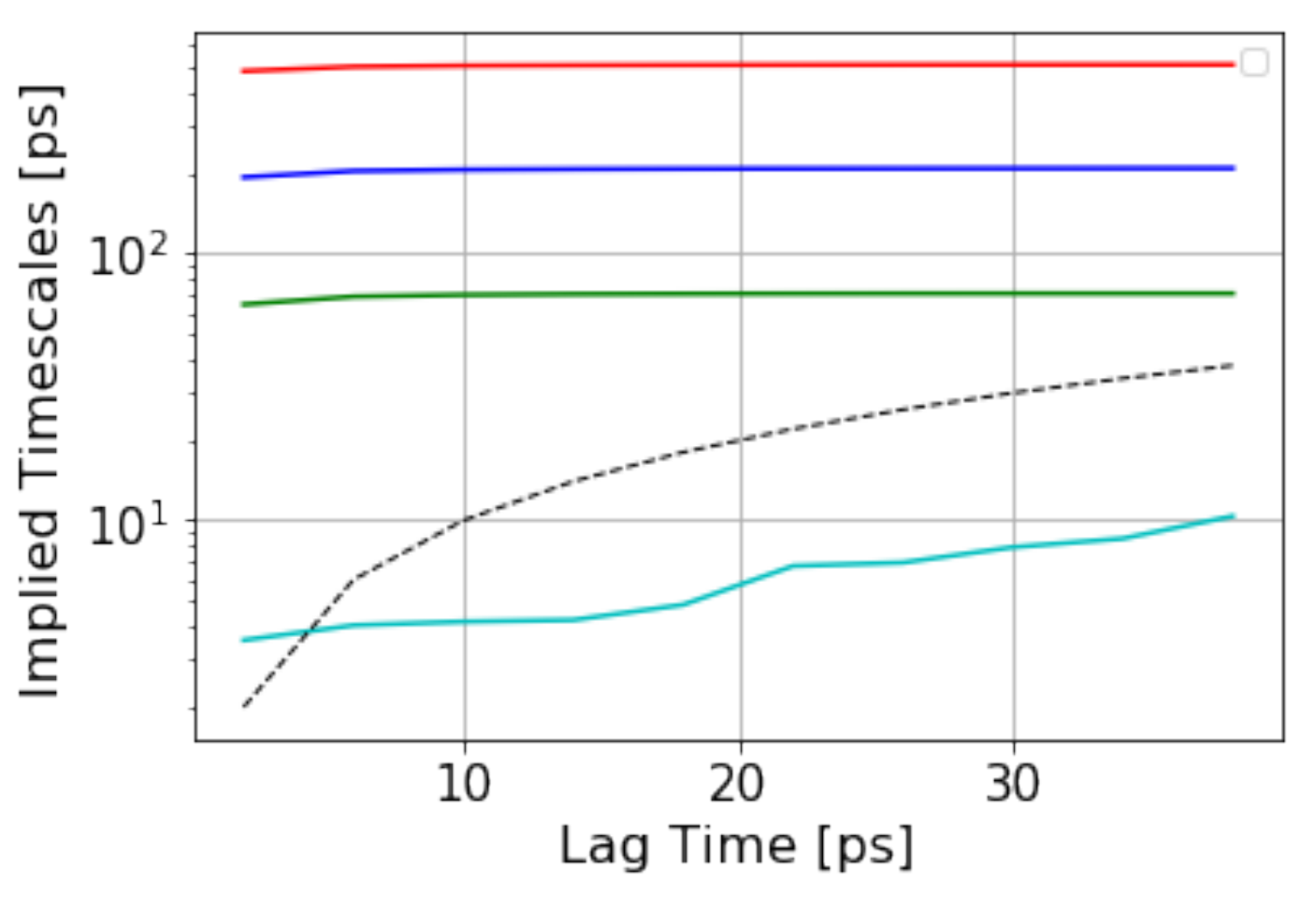}}
\caption{(a) Free energy map of dialanine in $\phi-\psi$ space. The 
black dashed lines correspond to the simple rectangular state decomposition 
that we sometimes use.  
(b) Relaxation timescales calculated as a function of lag time for 
dialanine in explicit water. The black line corresponds to $y=x$. Any 
relaxation process that is below this line cannot be estimated 
reliably since the process has already decayed.  
}
\label{ala-free}
\end{figure}
\begin{figure}
 \centering
 \includegraphics[width=0.6\columnwidth]{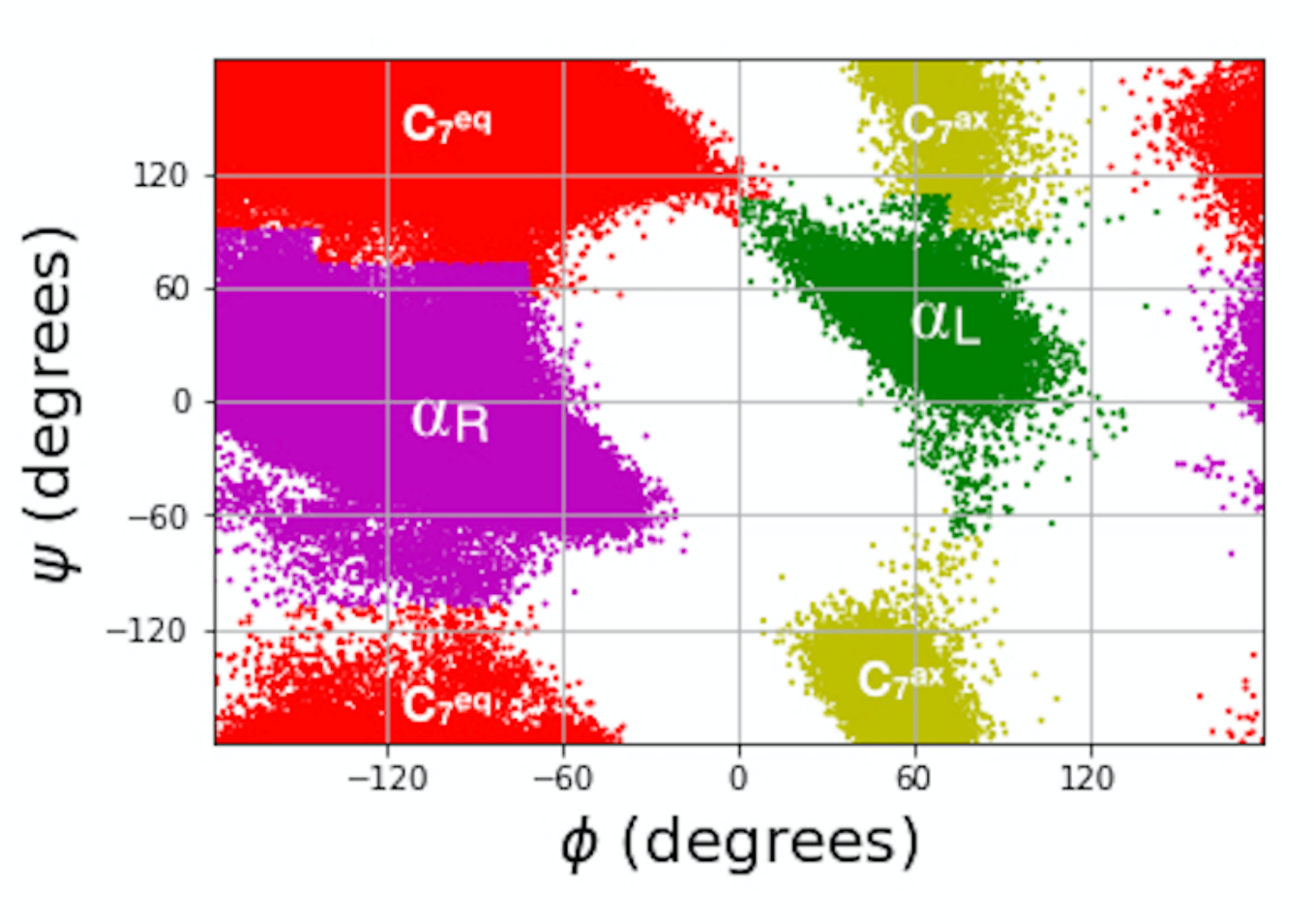}
 \caption{Four metastable states obtained from PCCA method for ananine 
 dipeptide in explicit water. The conformations are colored according 
 to the macrostates they belong to. The equilibrium populations of 
 the states are  $C_{7}^{ax}$ : 1.2\%, $\alpha_{L}$ : 4.7\%, $C_{7}^{eq}$ 
 : 42.9\%, $\alpha_{R}$ : 51.0\%} 
 \label{ala4}
\end{figure}
For MSM construction, the state space was discretized uniformly in 
$20\times 20$ bins, where each bin represents a single microstate.  
Based on the convergence of the implied timescales in Figure~\ref{ala-free:b}, 
we construct a MSM at a lag time of 20 ps.  The spectral analysis 
shows that there is a large timescale separation between the third 
and fourth relaxation timescales, indicating that there are four metastable 
states in the system; we thus employ PCCA to define four macrostates.  
The four states correspond to the $\alpha_{R}$, $C_{7}^{eq}$, $\alpha_{L}$ 
and $C_{7}^{ax}$ conformations of dialanine.  The PCCA macrostate 
boundaries obtained by crisp assignment of microstates to metastable 
sets (Eq. 2 in SI) are shown in Figure~\ref{ala4} along with 
their relative populations.  
\begin{figure}
\centering     %%% not \center
\includegraphics[width=1.1\columnwidth]{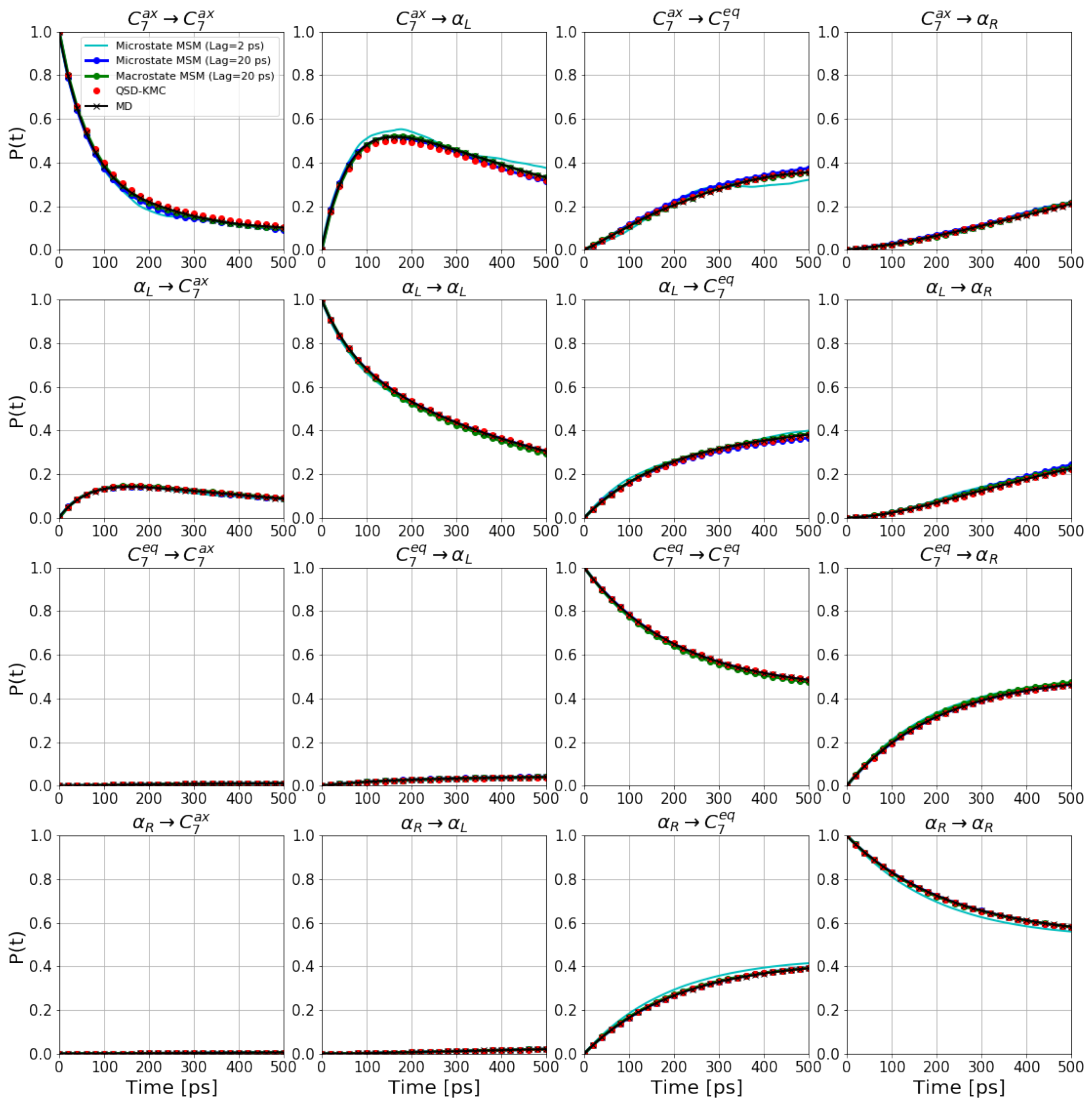}
\caption{State-to-state probability evolutions calculated from QSD-KMC 
and underlying MD trajectories for the four metastable PCCA states of  
dialanine. The dephasing times for all the states are estimated 
from the survival probability functions shown in Figure \textbf{S1}. For comparison, 
we also show the results from microstate MSM's constructed at lag times 
of 2 ps and 20 ps, and a macrsotate MSM constructed at a lag time of 20 ps.  
}
\label{ck-4-case-1}
\end{figure}
Based on the these boundaries, we generate a QSD-KMC trajectory and 
compare the resulting dynamics with MD projected onto these metastable 
states.  The dephasing times and the escape rates for the states are taken from the 
survival probability functions and Anderson-Darling test statistic plots 
shown in Figure \textbf{S1}.  
We can observe here that the Anderson-Darling test is very sensitive, giving 
dephasing times much longer than one might choose by eye.
The $C_{7}^{ax}$ state in Figure \textbf{S1(a)}, for which $\tau_d$=162 $ps$,
is a good example of this; after the first few $ps$, the nonexponentiality is
extremely subtle, but the Anderson-Darling test nonetheless detects it strongly, 
out to well beyond 100 $ps$.
\begin{figure}
\centering     %%% not \center
\subfigure[Initial Partitioning]{\label{ala-opt-4:a}\includegraphics[width=60mm]{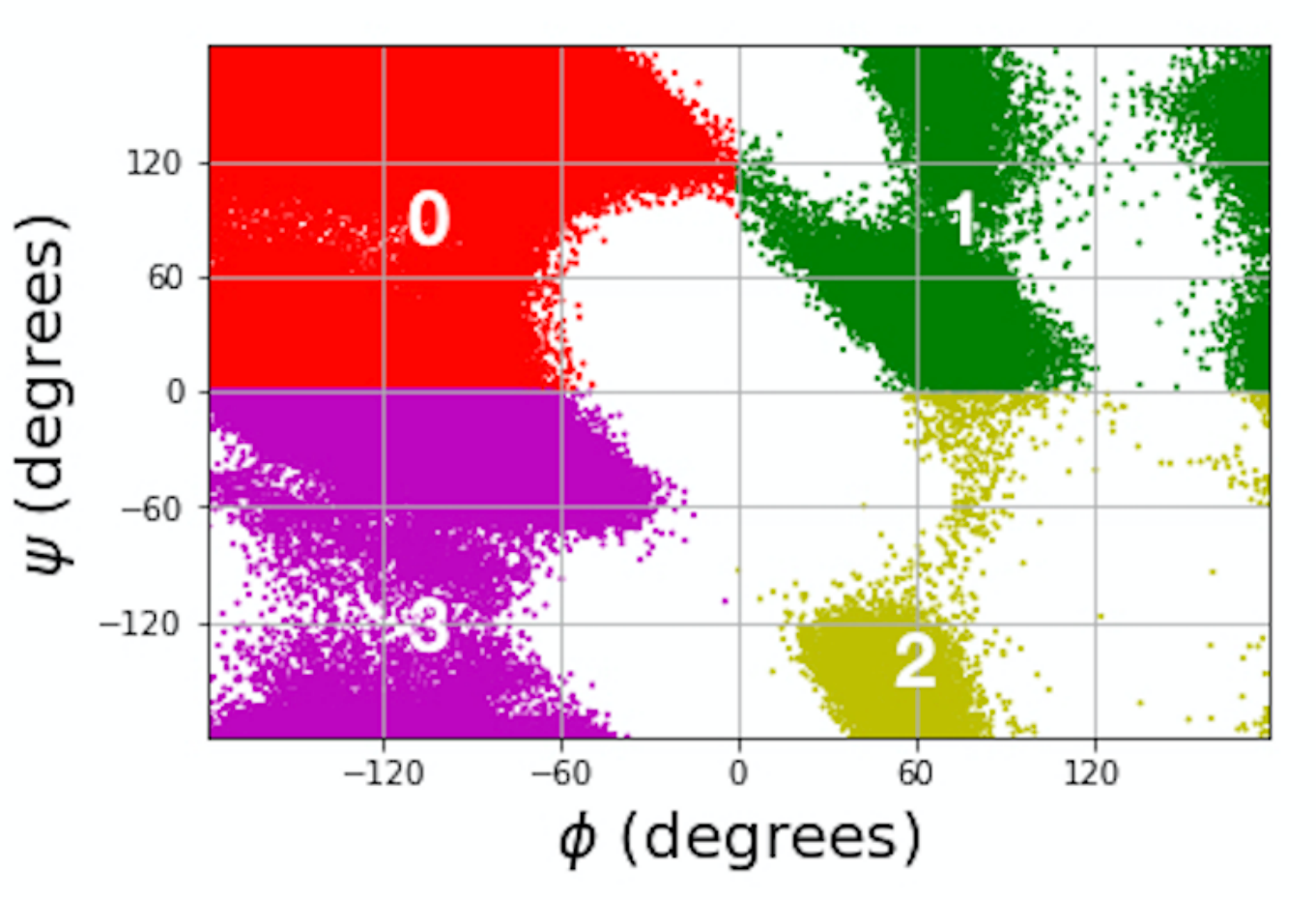}}
\subfigure[Final Partitioning]{\label{ala-opt-4:b}\includegraphics[width=60mm]{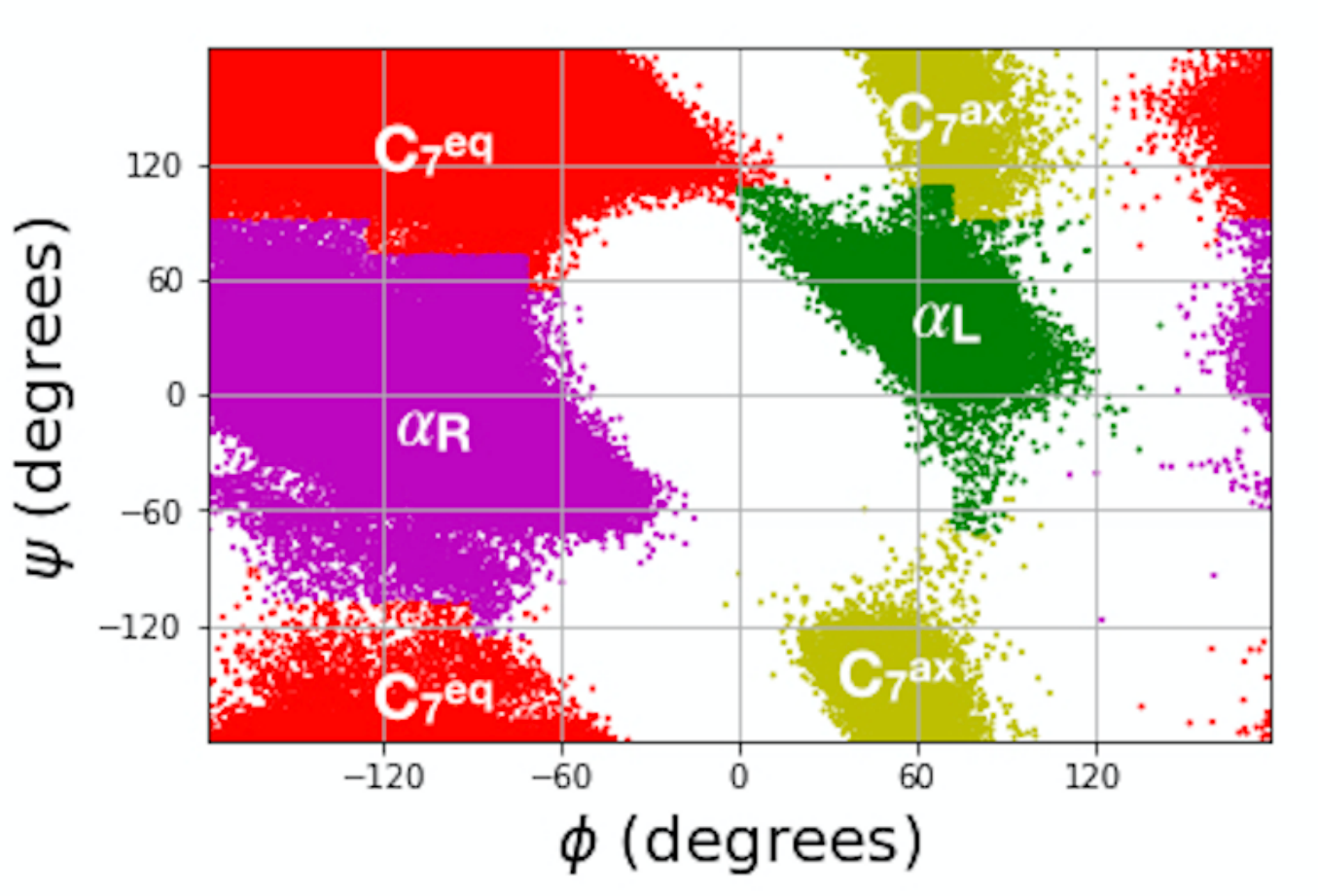}}
\caption{Partitioning of the $\phi-\psi$ space for dialanine in four states (a) Rectangular lumping of cells into four macrostates 
(b) Partitioning obtained by the application of the outside-time-based state optimization method.}
\label{ala-opt-4}
\end{figure}
We also compute the probability evolutions from a microstate MSM 
constructed at lag times of 2 $ps$ and 20 $ps$ as well as from a 
macrostate MSM constructed at a lag time of 20 $ps$. All these results 
are shown in Figure~\ref{ck-4-case-1}.  We see that the QSD-KMC results 
(red circles) are in very good agreement with the MD probability 
evolutions (solid black line).  It can also be seen that the results 
from different Markov models also agree quite well with the MD results, 
indicating the Markovian behavior of these PCCA macrostates.  
\begin{figure}
\centering     %%% not \center
\includegraphics[width=1.1\columnwidth]{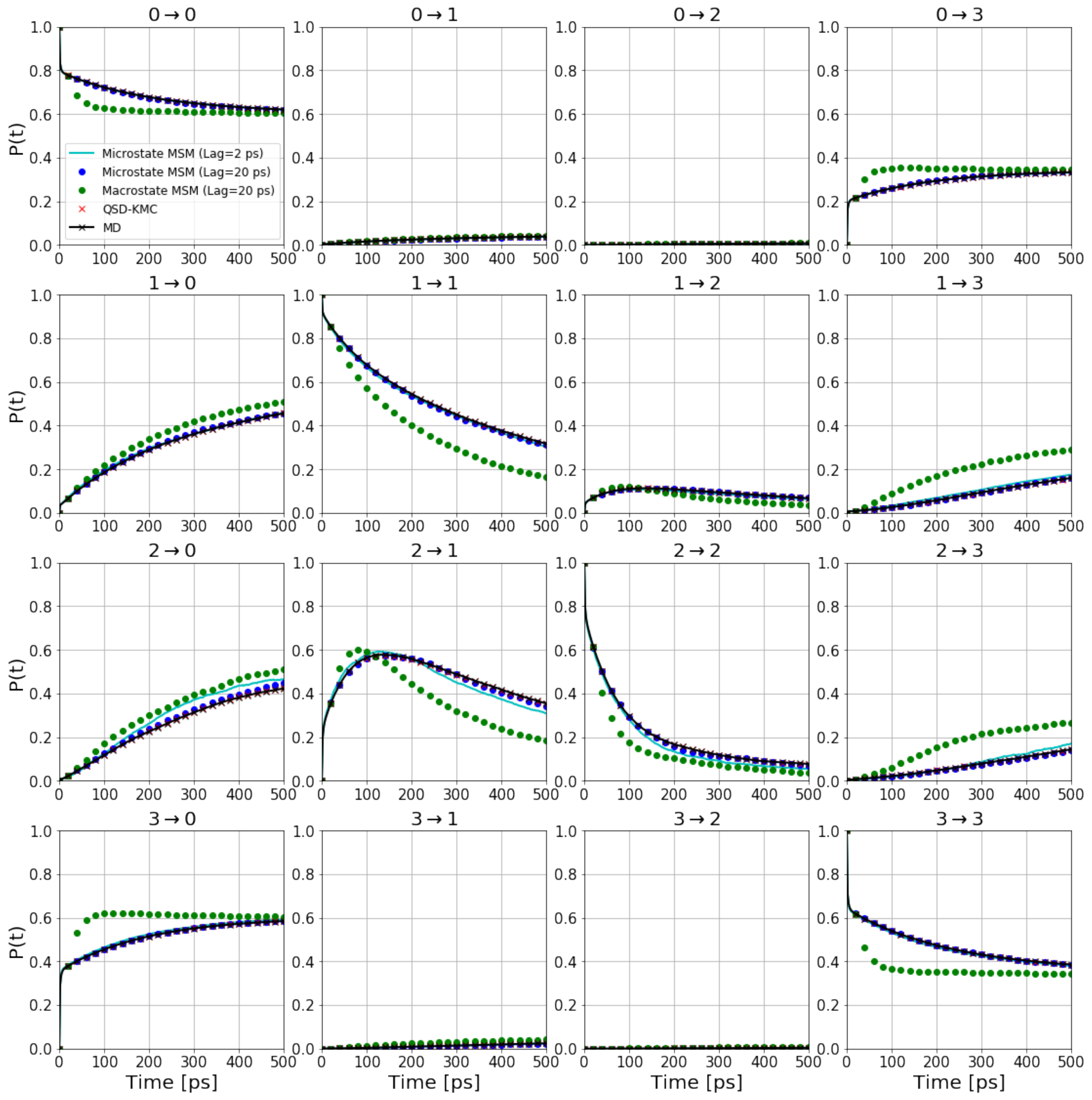}
\caption{State-to-state probability evolutions calculated from QSD-KMC 
and underlying MD trajectories for the simple rectangular states of 
dialanine.  The dephasing times for all the states are estimated 
from the survival probability functions shown in Figure \textbf{S2}. 
For comparison, we also show the results from microstate 
MSM's constucted at lag times of 2 ps and 20 ps, and a macrsotate 
MSM constructed at a lag time of 20 ps.  } 
\label{ck-4-case-2}
\end{figure}

To further demonstrate the robustness of our method, we construct 
an artificial set of states that ignores the physics of the system 
by taking a simple rectangular partition of 
the $\phi-\psi$ space
(State 
0: $\phi \in$ [-180, 0] and $\psi \in$ [0, 180], State 1: $\phi \in$ 
[0, 180] and $\psi \in$ [0, 180], State 2: $\phi \in$ [0, 180] and 
$\psi \in$ [-180, 0] and State 3: $\phi \in$ [-180, 0] and $\psi 
\in$ [-180, 0])). 
Figure~\ref{ala-opt-4:a} shows the states defined according to these boundaries.
We determine the dephasing times and escape rates for these states 
based on the survival probability plots shown in Figure \textbf{S2}.  
We also construct the various Markov state models as we did just above for the previous case. It 
can be seen from Figure~\ref{ck-4-case-2} that probability evolutions 
between different states calculated from QSD-KMC are in excellent agreement 
with the evolutions calculated from the underlying MD.  The 
results from microstate MSM also agree quite well with the probabilities 
computed in the underlying dynamics; however, for this rectangular set 
of states, the macrostate MSM (green circles) fails to capture the 
underlying dynamics even at long lag times.  This can be attributed 
to the highly non-Markovian behavior of the dynamics when the trajectory 
is mapped onto these rectangular macrostates.  In contrast, QSD-KMC 
gives a highly accurate description of the dynamics at every time resolution 
since it accounts for the correlated events. This demonstrates 
the effectiveness of the method to evolve a system from state to 
state, irrespective of the nature of the underlying dynamics.
\newline

\textbf{State Optimization}: We initiate the state-optimization process 
with rectangular lumping of 20$\times$20 bins into four macrostates. 
The initial state decomposition is shown in Figure~\ref{ala-opt-4:a}.  
The timescales obtained using these definitions of the states (dashed 
lines in Figure \textbf{S3}) deviate substantially from 
the timescales obtained using the PCCA states (solid lines in Figure \textbf{S3}).  
Again, the dephasing times are selected based on the survival probability 
functions shown in Figure \textbf{S2}, in accordance with 
the procedure described in Section~\ref{choice}.  
The application of the outside-time-based optimization procedure 
leads to the partitioning into four metastable states shown in Figure~\ref{ala-opt-4:b}.  
We see that the four states obtained have their boundaries at roughly 
the same location as those obtained via the PCCA method (Figure~\ref{ala4}).  
We also see, from Figure \textbf{S3}, that the implied timescales 
obtained from the transition matrix constructed between the optimized states 
are very similar to those from PCCA. As mentioned in Section~\ref{optimization}, we construct the survival probability functions 
for the four states obtained via the application of our method (shown 
in Figure \textbf{S4}) and repeat the state optimization procedure with this optimized set of states and the new
dephasing times (Figure \textbf{S4}). We find that the state boundaries remain unchanged. 
Thus, our QSD-KMC-based state optimization method can 
be used as an alternative to existing methods such as PCCA.  

\begin{figure}[H]
\centering     %%% not \center
\subfigure[Free-energy landscape of villin headpiece]{\label{vil-free:a}\includegraphics[width=65mm]{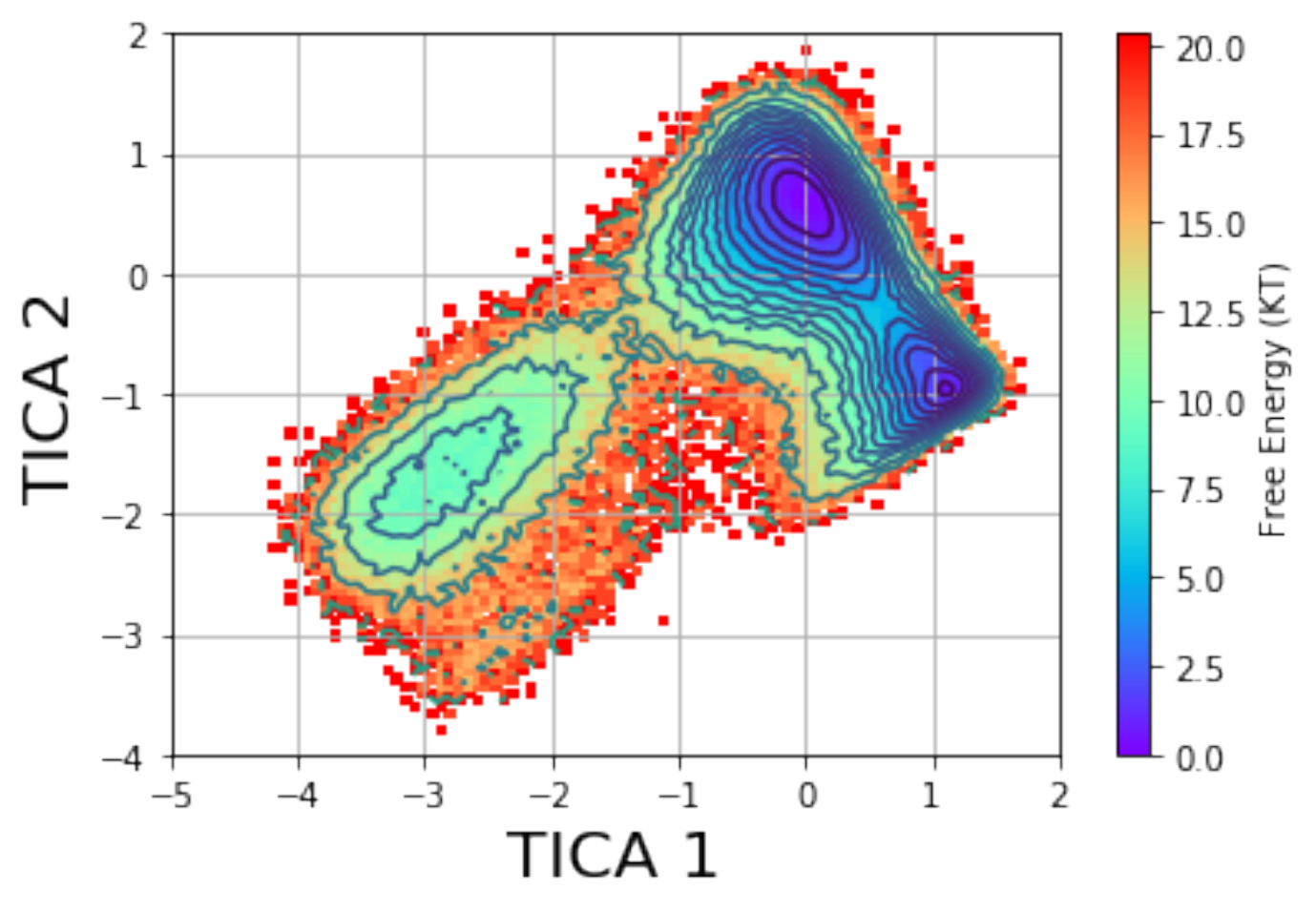}}
\subfigure[Relaxation timescales calculated as a function of lag time for villin headpiece]{\label{vil-free:b}\includegraphics[width=65mm]{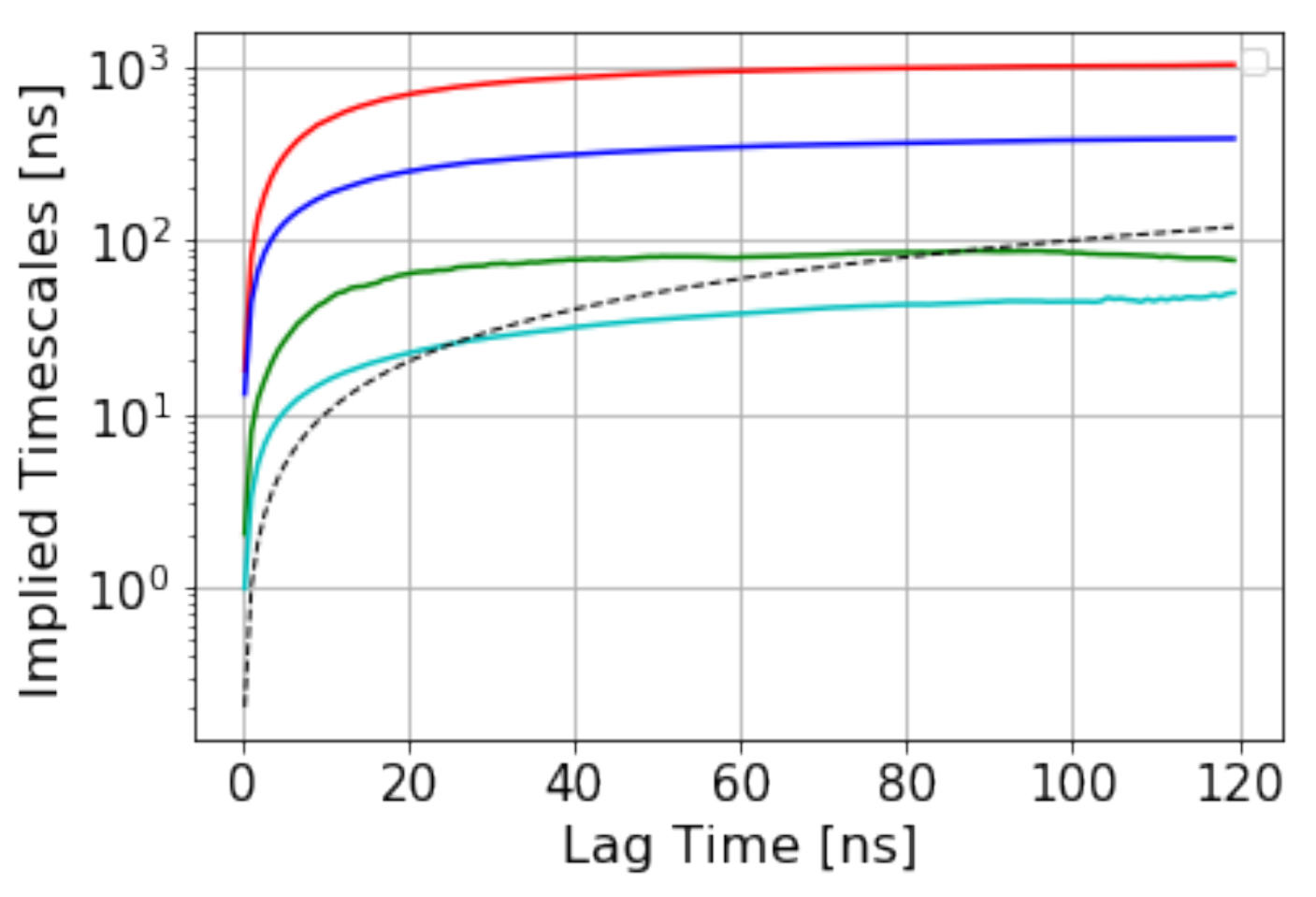}}
\caption{(a) Apparent free-energy landscape of villin headpiece along the first two components of TICA
(b) Relaxation timescales calculated as a function of lag time for villin headpiece. Black line corresponds to $y=x$. Any relaxation process 
which is below this line cannot be estimated reliably since the process has already decayed.
}
\label{villin-free}
\end{figure}

\noindent
\textbf{Villin headpiece}:  

We test the applicability of the QSD-KMC 
method on the folding of villin headpiece, a fast-folding protein. 
This has been a prototypical protein system for studying folding 
due its small size, simple secondary structure topology, and fast 
folding rate of a few microseconds. We considered a folding trajectory 
consisting of a 125 $\mu s$ MD simulation 
at 360 $K$ where the snapshots were saved at 0.2 $ns$ time intervals.
Multiple folding and unfolding events occur in this time. This simulation 
was performed by D.E. Shaw Research on the ANTON supercomputer~\cite{anton}; 
technical details of the simulation are reported in~\cite{folding}.  
Even though this is a single, long MD trajectory, the QSD-KMC model 
generated from it should be essentially the same as what we would 
obtain from a large set of short, Boltzmann-initiated trajectories 
summing to the same total time.
 \begin{figure}[H]
\centering
\includegraphics[width=0.65\columnwidth]{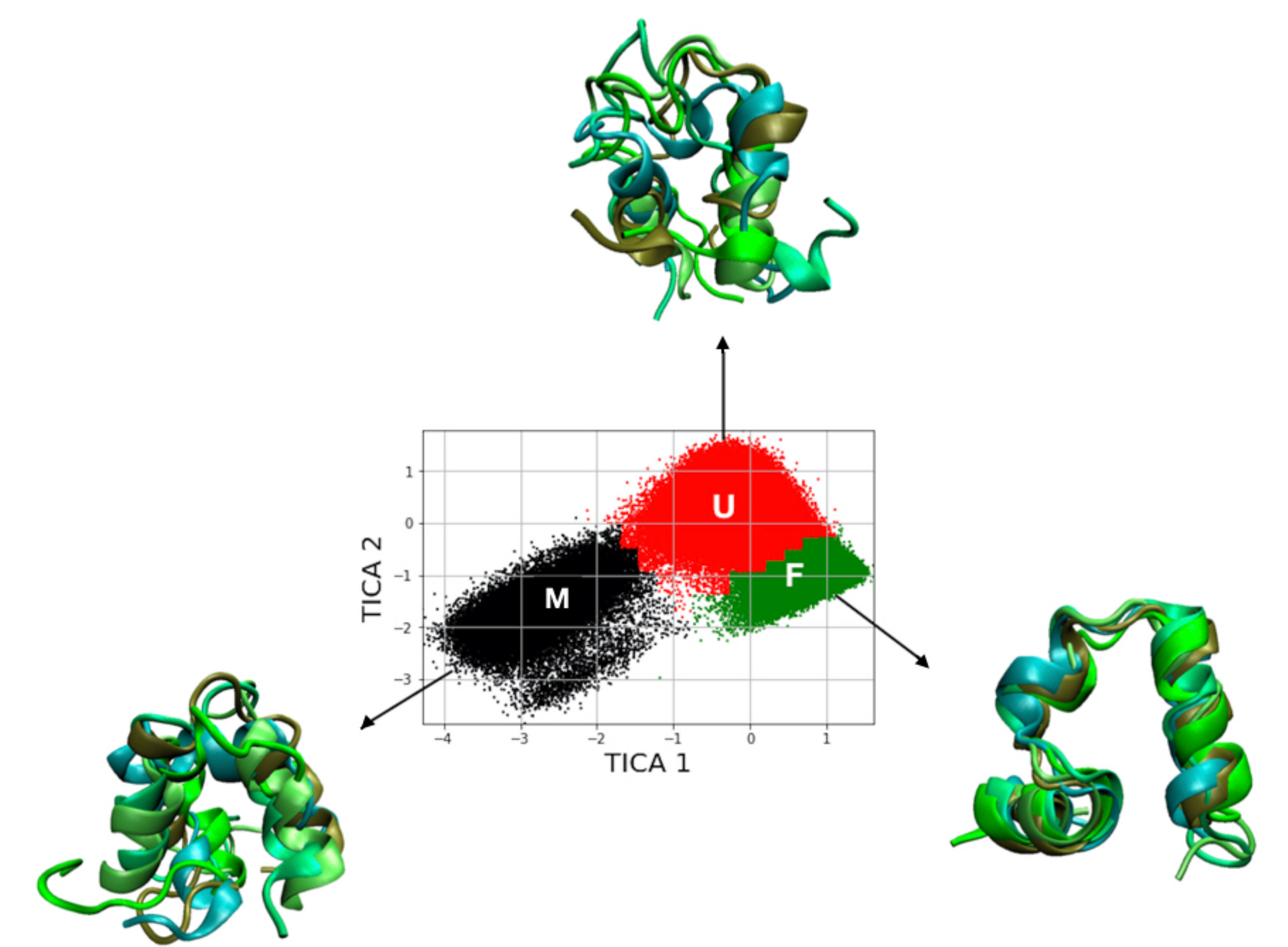}
\caption{Three metastable states obtained from PCCA method for villin headpiece. 
These states can be characterized as Folded (F), Unfolded (U) and 
Misfolded (M). The conformations are colored according to the macrostates 
they belong to. For every state, we have shown a representative structure. 
The equilibrium populations of the states are:  M : 5.78\%, F : 23.12\%, 
U : 71.10\%} 
\label{pcca-vil-state-3}
\end{figure}
We employ time-lagged independent component analysis (TICA)~\cite{tica} 
at a lag time of 50 $ns$ to obtain the two slowest degrees of freedom. 
As a feature set, we select the backbone and side chain torsion angles 
in the protein.  
Figure~\ref{vil-free:a} shows the free-energy map in the TICA coordinate 
space. This space is discretized into 675 microstates by uniform 
discretization into $25 \times 25$ bins.  
The implied timescales plot (Figure~\ref{vil-free:b}) shows 
that the timescales have reached a plateau at 40 $ns$, so we construct 
a MSM at this lag time.  The spectral analysis shows that there is 
a large timescale separation between the second and third relaxation 
timescale.  We therefore generate three metastable states using PCCA 
as shown in Figure~\ref{pcca-vil-state-3}. We randomly draw 200 structures 
out of these three macrostates to get a sense of these states with 
respect to folding. Based on secondary structure content and native 
contacts analysis (Figure \textbf{S5} and \textbf{S6} of the supporting information), 
we characterize these metastable states as Folded (F), Unfolded (U) 
and Misfolded (M)  (Figure~\ref{pcca-vil-state-3}).  The equilibrium 
populations of the states are:  M : 5.78\%, F : 23.12\%, U : 71.10\% 
which is indicative of the temperature at which the simulation was 
carried out. The Unfolded state,U, contains a large number of microstates,
as one would expect from a folding funnel perspective with high configurational 
entropy. Conformations from this basin show helix 1 (residues 44-51) 
and helix 3 (residues 55-60) are partially formed whereas no significant 
content is seen for helix 2 (residues 63-73). The Misfolded state, 
M, shows a well-folded helix 3 that extends onto the N-terminal region 
to encompass some part of helix 2 with partial folding of helix 1. 
Compared to the folded state, a critical loop region between 
helix 2 and helix 3 is missed and a coil region is seen at 
the end of helix 3. The Folded state, F, has a  compact tertiary 
structure where helix 2 is formed along with helix 1 and helix 3.  
\begin{figure}[H]
\centering     %%% not \center
\includegraphics[width=1.1\columnwidth]{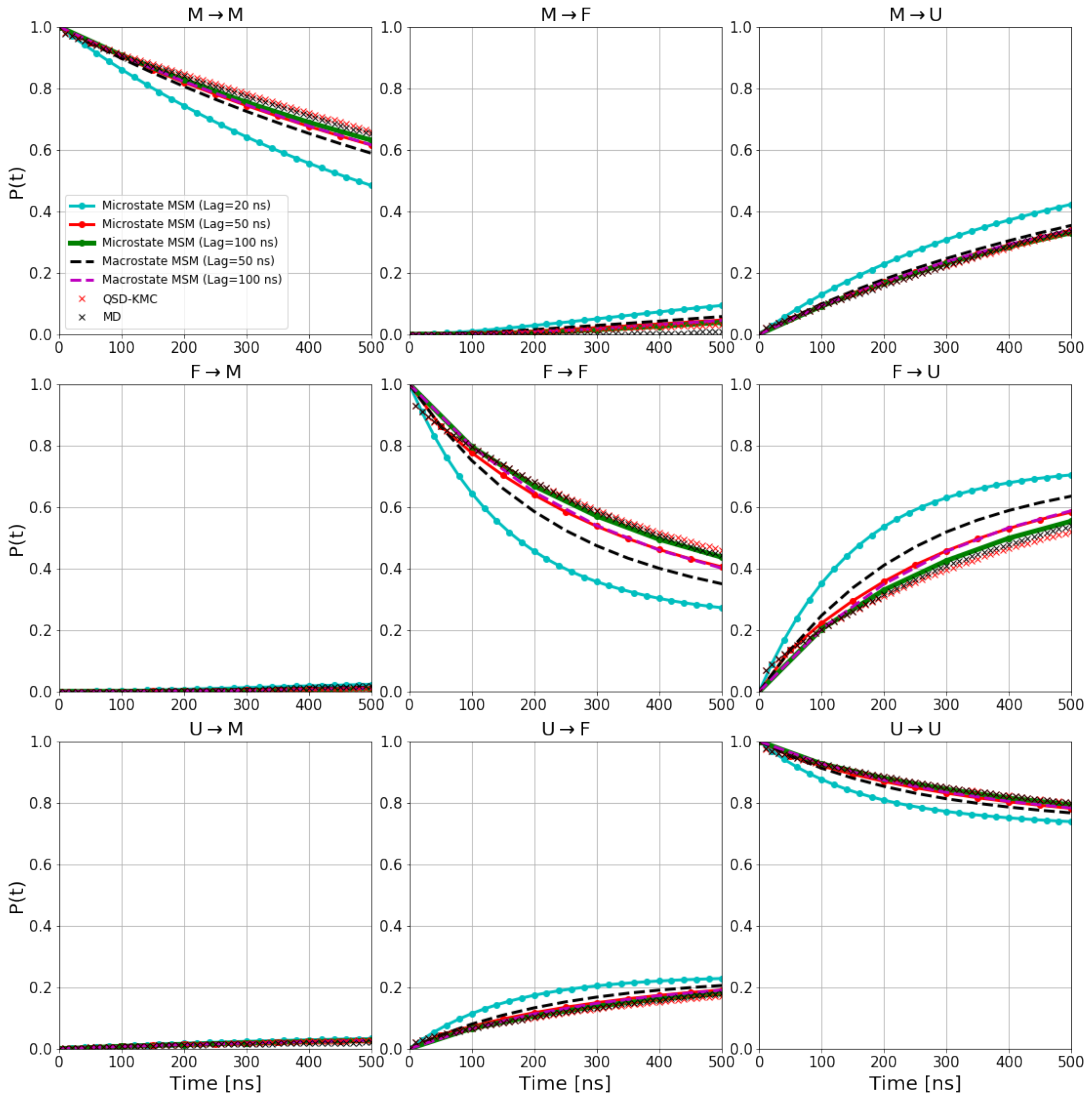}
\caption{State-to-state probability evolutions calculated from QSD-KMC 
and underlying MD trajectories for the three PCCA metastable states 
of villin headpiece.  The dephasing times for all the states are 
estimated from the survival probability functions shown in Figure~\ref{survival-vil-1}.  
For comparison, we also show the results from microstate MSM's constucted 
at lag times of 20 ns, 50 ns and 100 ns, and a macrostate MSM constructed 
at a lag time of 50 ns and 100 ns.} 
\label{ck-villin-case-1}
\end{figure}
We estimate the dephasing times ($\tau_d^M$=81.0 $ns$, $\tau_d^F$=62.6 $ns$, $\tau_d^U$=30.4 $ns$)
and total escape rates for these states from the survival 
probability function and Anderson-Darling test statistic plots shown in Figure~\ref{survival-vil-1}. 
We compute the probability 
evolutions using QSD-KMC, microstate MSM at different lag times (20 
$ns$, 50 $ns$ and $100$ ns), and macrostate MSM at a lag time of 50 
$ns$ and 100 $ns$.  It can be seen in Figure~\ref{ck-villin-case-1} 
that the predictions from MSM at a lag time of 100 $ns$ agree with 
the MD results to a satisfactory degree of accuracy; however, the 
probability evolutions calculated from Markov models constructed 
at time resolution less than 50 $ns$ deviate from the MD results.  
The QSD-KMC trajectory, however, gives an accurate description of 
the dynamics even at the finest time resolution of 0.2 $ns$. 
\begin{figure}[H]
\centering     %%% not \center
\subfigure[Initial Partitioning]{\label{vil-opt-3:a}\includegraphics[width=59mm]{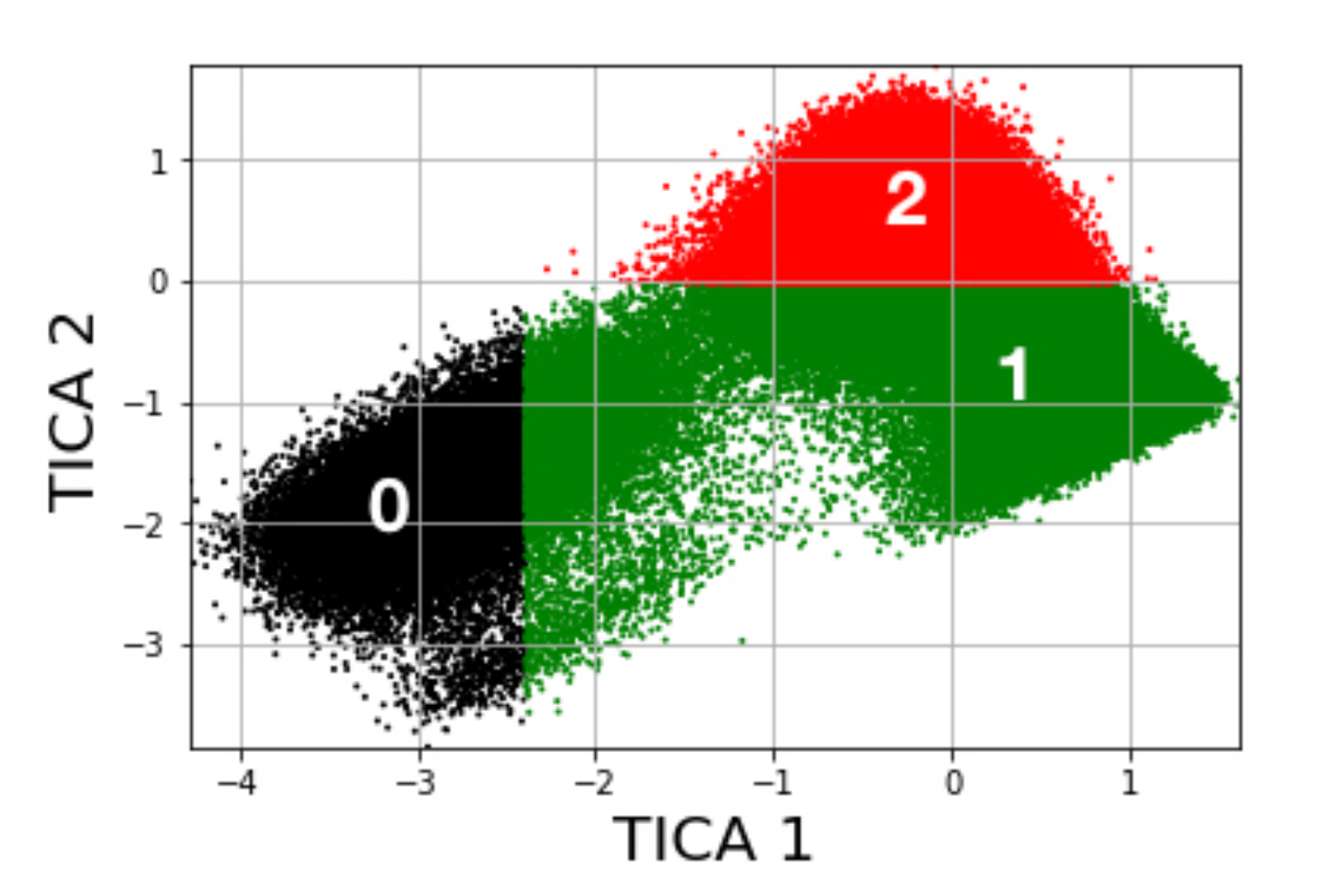}}
\subfigure[Final Partitioning]{\label{vil-opt-3:b}\includegraphics[width=57mm]{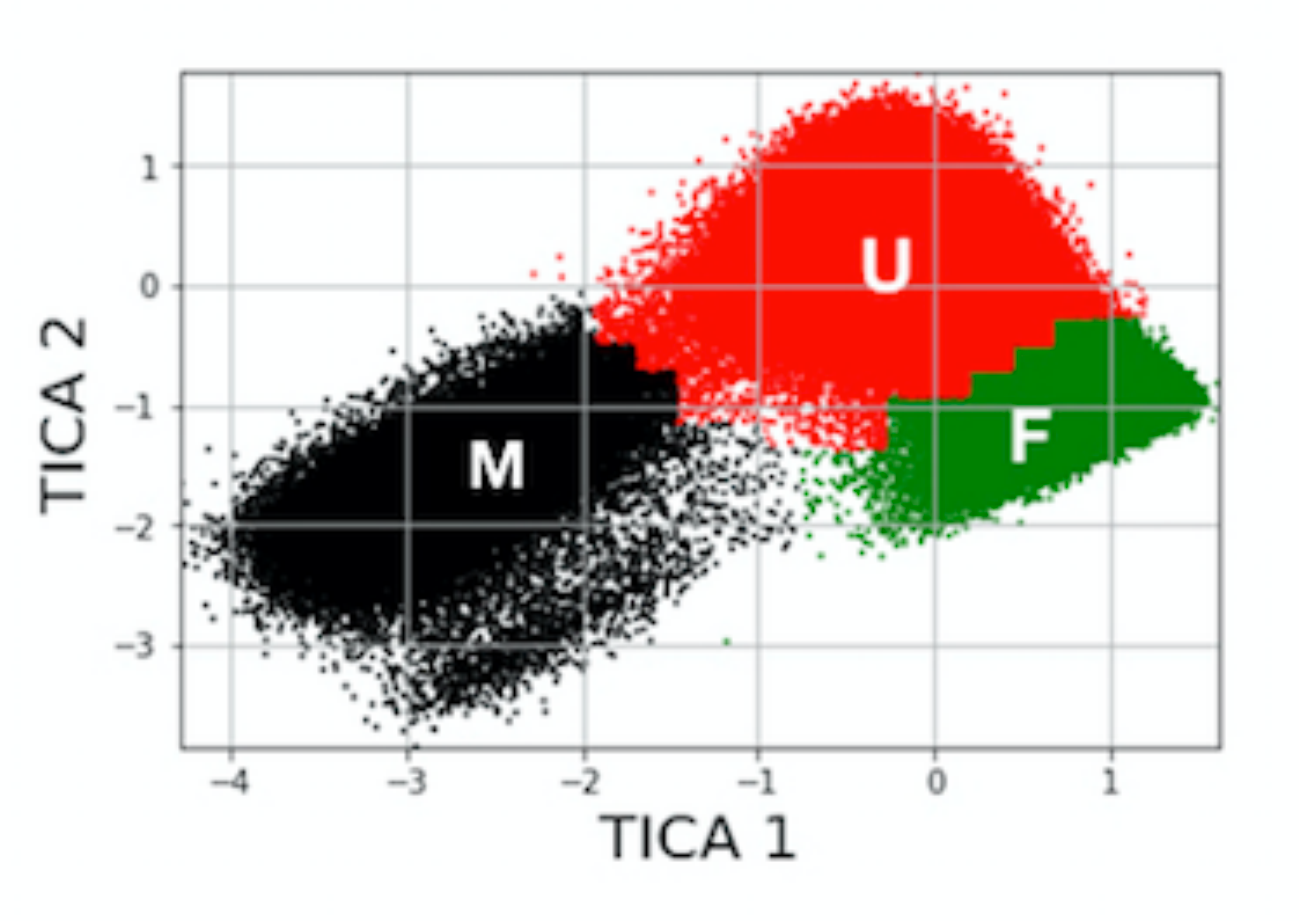}}
\caption{Partitioning of the TICA-based coordinate space for villin 
headpiece into three states (a) Rectangular lumping of cells.  
(b) Partitioning obtained by the application of the outside-time-based 
state optimization method.} 
\label{vil-opt-3}
\end{figure}
We also test the effectiveness of our approach where an unphysical rectangular 
state space has been specified, as shown in Figure~\ref{vil-opt-3:a}.  
Based on the survival probability functions and Anderson Darling test statistic plots shown in Figure \textbf{S7}, 
we select a conservative value of $\tau_{d}$ for states 0, 1 and 
2 as 5.4 $ns$, 62.0 $ns$ and 24.4 $ns$ respectively.  Figure~\ref{ck-villin-case-2} 
shows that the QSD-KMC results are in excellent agreement with the probability 
evolutions computed from the underlying MD trajectories, for this 
rectangular set of macrostates.  The results from the microstate 
MSM are similar to the dialanine case; i.e. the predictions agree 
with the underlying MD results if a long lag time is chosen. If a 
MSM is constructed directly over this set of macrostates, 
the probability evolution results deviate quite a lot from MD even 
at a lag time of 100 $ns$.  Thus, if one is interested in a fully 
resolved and continuous representation of the dynamics with high 
accuracy, QSD-KMC is a good choice.

\textbf{State Optimization}: As an initial choice of a three state 
partitioning, we perform the rectangular state decomposition shown 
in Figure~\ref{vil-opt-3:a}.  Figure \textbf{S8} shows 
that the implied timescales obtained from the spectral 
analysis of the transition matrix constructed over these initial 
states (dashed lines) strongly deviate from timescales obtained from 
crisp PCCA states (solid lines).  
Application of the outside-time-based state 
optimization method produces the metastable set of states shown in 
Figure~\ref{vil-opt-3:b}; the state boundaries obtained are at 
locations that are very similar to the ones obtained from PCCA (Figure~\ref{pcca-vil-state-3}). 
The comparison of the relaxation timescales obtained between the 
optimized set of states and PCCA states (Figure \textbf{S8}) 
demonstrate the potency of the method to optimize the boundaries 
for a given set of macrostates.  
\begin{figure}[H]
\centering     %%% not \center
\includegraphics[width=1.1\columnwidth]{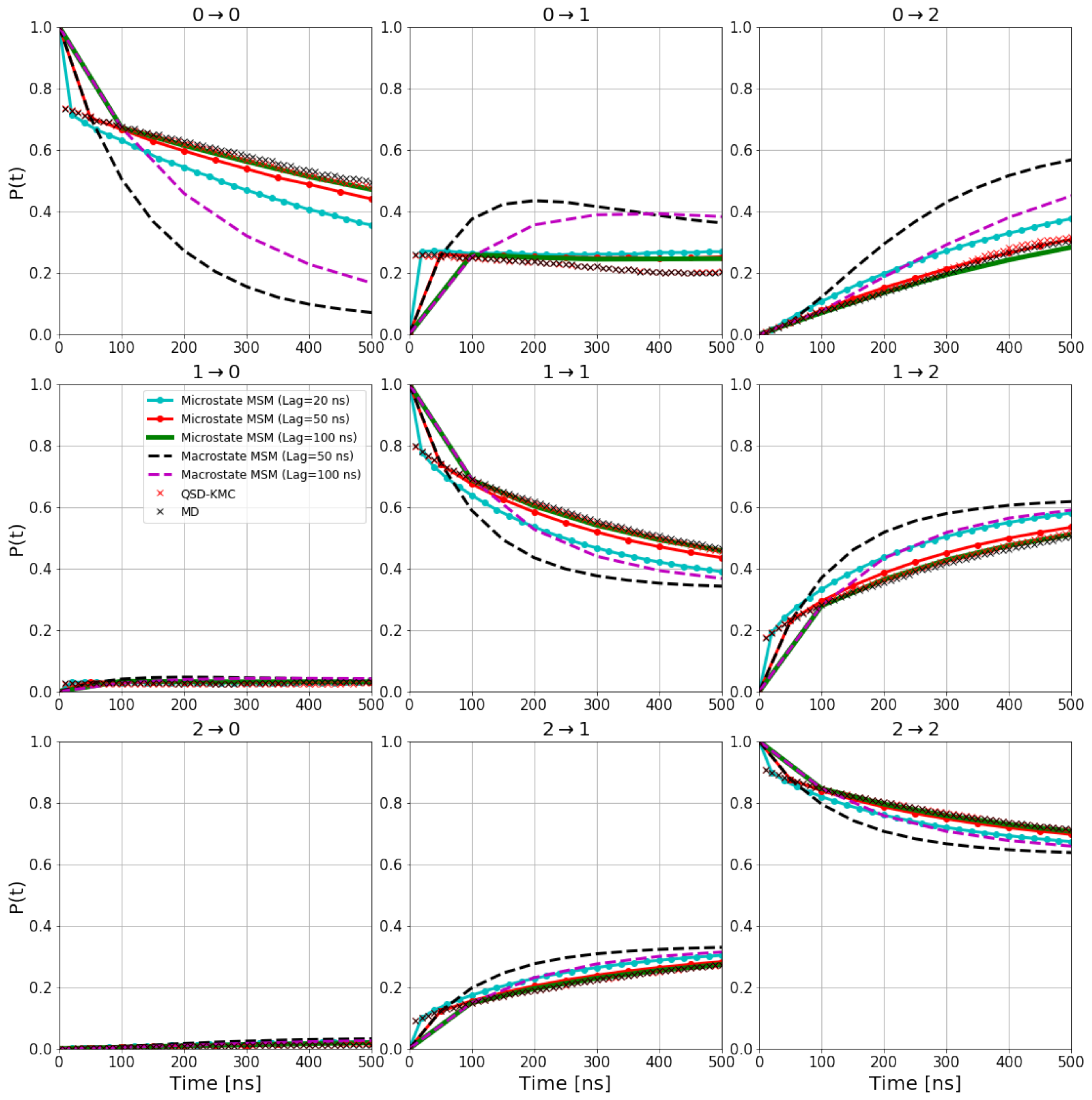} 
\caption{State-to-state probability evolutions calculated from QSD-KMC 
and underlying MD trajectories for the simple rectangular states of 
villin headpiece.  The dephasing times for all the states are estimated 
from the survival probability functions shown in Figure \textbf{S7}.  
For comparison, we also show the results from microstate MSM's constucted 
at lag times of 20 ns, 50 ns and 100 ns, and a macrostate MSM constructed 
at a lag time of 50 ns and 100 ns.} 
\label{ck-villin-case-2}
\end{figure}

\section{Conclusions}
We have introduced the QSD-KMC method, a new computational approach 
that generates accurate state-to-state evolution for complex biological 
systems using information from a set of relatively short MD trajectories.
The QSD-KMC model consists of a set of escape rates out of different 
states and a database of representative instantiations for escape from those states,
consisting of the states that the system passes through and the state it finally settles into.  
QSD-KMC utilizes the concept of the quasi-stationary distribution 
to prepare a memoryless distribution within a state, and then properly 
accounts for correlated events that may occur upon exit from that 
state. 
Although achieving the QSD places constraints on the short trajectories 
that may increase the length of these trajectories compared to typical 
MSM trajectories, a significant benefit is that QSD-KMC generates 
arbitrarily accurate dynamics even when the states are defined in 
a way that does not exhibit ideal Markovian behavior.  
Furthermore, the theoretical concepts of QSD-KMC can be employed 
to design a Monte Carlo approach that optimizes the state boundaries 
starting from an arbitrary set of macrostates. 
We have demonstrated the important concepts of the method on two one-dimensional 
model systems.  We then applied the method to two different biological 
systems and showed that QSD-KMC gives more accurate results than 
the conventional Markov State Model.  Indeed, even with an intentionally 
nonphysical choice of the macrostate definitions, QSD-KMC gives results 
fully faithful to the underlying MD it is built on. Thus, QSD-KMC 
is a powerful approach that can be employed to understand the behavior on long timescales 
in complex biological systems where the complicated energy landscape 
prohibits a clear definition of Markovian states.  

\section{Supplementary Material}

See supplementary material for the technical details and additional figures. 

\section{Acknowledgements}

This work has been supported in part by the Joint Design of Advanced Computing Solutions for Cancer (JDACS4C) program 
established by the U.S. Department of Energy (DOE) and the National Cancer Institute (NCI) of the National Institutes 
of Health. This work was performed under the auspices of the U.S. Department of Energy by Argonne National Laboratory 
under Contract DE-AC02-06-CH11357, Lawrence Livermore National Laboratory under Contract DE-AC52-07NA27344, Los Alamos National Laboratory 
under Contract DE-AC5206NA25396, Oak Ridge National Laboratory under Contract DE-AC05-00OR22725, and Frederick National Laboratory for Cancer
Research under Contract HHSN261200800001E. We thank Prof. Angel E. Garcia and Dr. Danny Perez for helpful discussions. 
We thank Dr. Srirupa Chakraborty,  Dr. Timothy Travers and Dr. Cesar A. Lopez for helping us 
with the simulations and analysis of dialanine and villin headpiece. We also thank D.E. Shaw Research for providing the villin headpiece trajectory.
Computing resources were made available by LANL Institutional Computing.

\end{document}